\documentclass{aastex62}
\usepackage{amsmath,amsthm}
\received{*}
\revised{*}
\accepted{*}

\submitjournal{ApJS}
 
\shorttitle{thermal equilibrium mass loss in binaries}
\shortauthors{Ge et al.}

\begin{document}

\title{The thermal equilibrium mass loss model and its applications in binary evolution}
 
\correspondingauthor{Hongwei Ge; Ronald F Webbink}
\email{gehw@ynao.ac.cn; rwebbink@illinois.edu}

\author{Hongwei Ge}
\affiliation{Yunnan Observatories, Chinese Academy of Sciences, Kunming 650216, PR China}
\affiliation{Key Laboratory for the Structure and Evolution of Celestial Objects, Chinese Academy of Sciences, Kunming 650216, PR China}
\affiliation{Center for Astronomical Mega-Science, Chinese Academy of Sciences, Beijing 100012, PR China}
\affiliation{University of Chinese Academy of Sciences, Beijing 100049, PR China}

\author{Ronald F Webbink}
\affiliation{University of Illinois at Urbana-Champaign, 1002 W Green St, Urbana, 61801, USA}

\author{Zhanwen Han}
\affiliation{Yunnan Observatories, Chinese Academy of Sciences, Kunming 650216, PR China}
\affiliation{Key Laboratory for the Structure and Evolution of Celestial Objects, Chinese Academy of Sciences, Kunming 650216, PR China}
\affiliation{Center for Astronomical Mega-Science, Chinese Academy of Sciences, Beijing 100012, PR China}
\affiliation{University of Chinese Academy of Sciences, Beijing 100049, PR China}

\begin{abstract}

Binary evolution is indispensable in stellar evolution to understand the formation and evolution of most peculiar and energetic objects, such as binary compact objects, Type Ia supernovae, X-ray binaries, cataclysmic variables, blue stragglers, hot subdwarfs, and central binaries in planetary nebulae. Mass transfer in binary stars can change the evolutionary path and fate of the corresponding objects relative to what is expected from single stellar evolution. What is the critical mass ratio at which unstable mass transfer occurs is an unsolved fundamental problem in binary evolution. To resolve this issue, we construct the thermal equilibrium mass loss model and derive critical mass ratios for both thermal timescale mass transfer and unstable mass transfer, the latter of which occurs when the outer Lagrangian point, ${\rm L_2}$, is overfilled. Using several $3.2~M_\odot$ stellar models as examples, we study the stellar response to thermal equilibrium mass loss and present the thresholds for thermal timescale mass transfer. We study the possible mass transfer channels of binary systems containing a $3.2~M_\odot$ donor star, taking into account thermal timescale mass transfer, unstable mass transfer through ${\rm L_2}$, and dynamical timescale mass transfer. We repeat this simulation for a grid of donor stars with different masses (from 0.1 to 100 $M_\odot$ with Z = 0.02) and at different evolutionary stages, and present our results. The results show that unstable mass transfer due to the overfilling of the outer Lagrangian point may also play an essential role in the formation of common envelopes for late red giant branch and asymptotic giant branch donors.

\end{abstract}


\section{INTRODUCTION}
\label{intro}

Stellar structure and evolution theory is the cornerstone of astrophysics. A detailed understanding of the structure and evolution of stars requires substantial knowledge about the stellar interiors \citep{carr96} and external interactions. For example, the interaction between components in a close binary system changes the evolution path and the fate of a component from what it would have been as a single star \citep{eggl06}. As part of the stellar evolution theory, binary evolution helps to explain lots of mysterious phenomena in observed stellar objects. For example, the Algol paradox \citep{batt73}, in which the less massive star in a binary system is already a subgiant while the star with much greater mass is still on the main-sequence, is resolved with mass transfer between the two components. Binary compact objects, such as double black holes, double neutron stars, and double white dwarfs, have been studied for decades \citep[e.g.,][]{post14}. They are arguably the most interesting and energetic observed objects, and are the main sources of gravitational waves through the merging process \citep{abad10,abbo16,abbo17}. These objects are mainly from binary evolution. Type Ia supernovae (SNe Ia) have been widely used as cosmological distance indicators, leading to the discovery of the acceleration of cosmic expansion. They mainly result from binary interactions \citep{meng17,wang18} too. As a matter of fact, almost all kinds of important classes of stellar objects are related to binary evolution. These include X-ray binaries \citep{pods02}, cataclysmic variables \citep{howe01}, blue stragglers \citep{chen08,gell11}, hot subdwarfs \citep{saio00,han02}, central binary stars in planetary nebulae \citep{jone17}, etc. Therefore, binary evolution is indispensable in stellar evolution and for understanding the whole evolution picture of related stellar objects.

Half of the stars are in binaries, and the binary fraction is higher for massive stars \citep{lang12,moe17}. This explains partly the importance of binary interactions. There are various interactions in binary systems, e.g. mass transfer, accretion, tidal synchronization, magnetic braking, and gravitational wave radiation etc. But the fundamental process in binary systems is mass transfer.

Mass transfer is what we mainly focus on here. It is generally assumed that mass transfer occurs when a companion fills its Roche lobe either due to its stellar evolutionary expansion or the orbital shrinking owing to any angular momentum loss mechanism. Based on the Roche geometry, Roche-lobe overflow (RLOF)  is widely used to describe the mass transfer process in binary evolution. The primary picture is that mass from the lobe-filling star (donor) is transferred to its companion (accretor) through the inner-Lagrangian point, ${\rm L_1}$. The mass transfer process is complicated due to the uncertainty of its mass transfer rate. The mass transfer rate depends on the initial mass (with given metallicity $Z$), $M_{\rm i}$, initial mass ratio, $q_{\rm i}$, initial orbital period, $P_{\rm orb}^{\rm i}$, and orbital change due to systematical mass loss and angular momentum loss. The mass transfer rate varies and adjusts according to the local state including the density and sound speed of the donor near ${\rm L_1}$, and the detailed systematic mass-loss and angular momentum-loss processes in a given binary system. If the mass-transfer rate is fast, the donor star first tries to readjust its structure to recover its equilibrium state. Because hydrostatic readjustment happens on the star's dynamical timescale, which is much shorter than the Kelvin-Helmholtz timescale, the response of the donor to rapid mass loss is almost adiabatic \citep{webb85,hjel87,ge10,delo10}. We have built the adiabatic mass loss model, which has a frozen entropy profile with mass in the star, to study the adiabatic response of the donor star to very rapid mass transfer. We have derived threshold conditions for dynamical timescale mass transfer \citep{ge10,ge15,ge19}, which can be used as the physical input for binary population synthesis studies. On the contrary, if the mass transfer rate is slow, on the timescale of nuclear evolutionary expansion, the star is in both thermal equilibrium and hydrostatic equilibrium. If the mass transfer rate is {\it moderate}, on the thermal timescale ($\tau_{\rm KH}$), the star loses its thermal equilibrium initially and tries to reach a new thermal equilibrium state. 

The thermal timescale mass transfer process lies in two crucial positions in binary evolution. First, donor stars on the Main Sequence (MS) or Hertzsprung Gap (HG) with a mass larger than about $1.3 M_{\odot}$ have a radiative envelope. There is a delayed dynamical timescale mass transfer if the convective core is exposed, and the initial mass ratio is large enough \citep{ge10}. So the initial thermal timescale mass transfer of these radiative donor stars calls our attention. A contact phase may be inevitable if the mass ratio is large enough, even for thermal timescale mass transfer because the companion star can not adjust the mass transferred from the donor star to thermal equilibrium, or the radius of the donor becomes more extensive than its outer Lagrangian point, ${\rm L_2}$.  Secondly, the thermal timescale, $\tau_{\rm KH}$, of a red giant branch (RGB) or asymptotic giant branch (AGB) star is so short ($\tau_{\rm KH}$ is around or less than $10^2$ yr) and is comparable with its dynamical timescale \citep{ge19}. Hence, the rapid mass transfer, even on a thermal timescale, for donor stars on late RGB/AGB is likely to result in the formation of a common envelope (CE), for the reasons of fast enough mass transfer or overfilling of the outer Lagrangian point, ${\rm L_2}$.

To get the critical threshold condition for thermal timescale mass transfer, we build thermal equilibrium mass loss models following the pioneering work by \citet{hjel89a,hjel89b}. We assume the time derivative of the specific entropy, ${\rm d}s/{\rm d}t$, is frozen with mass (explained in section 2). This means the stellar thermal relaxation (by which the star regains its thermal equilibrium) ability is frozen in a mass coordinate. Based on the thermal equilibrium mass loss model, we study the stellar responses and derive the critical mass ratios for thermal timescale mass transfer. Instead of merely studying the donor star's radius response to mass loss, we compare the donor star's inner radius ($R_{\rm KH}$; as defined in section 4.1) response with its Roche lobe radius ($R_{\rm L}$) response to get the critical mass ratio for thermal timescale mass transfer, $q_{\rm th}$. {\it The matter between the inner radius, $R_{\rm KH}$, and surface radius, $R$, is assumed to be an isothermal flow, which transfers mass to its companion through the ${\rm L_1}$ point and can accelerate on a thermal timescale to a mass transfer rate\footnote{The index ${\rm i}$ refers to the initial quantity when the donor overfills its Roche lobe in the formula.}, $\dot{M}_{\rm KH} = -M_{\rm i}/\tau^{\rm i}_{\rm KH}$}. Additionally, we compare the radius, $R$, to its Roche-lobe radius, $R_{\rm L_2}$, at outer Lagrangian point ${\rm L_2}$ during thermal equilibrium mass loss to find the critical mass ratio $q_{\rm L_2}$ using $\zeta_{\rm eq} \equiv ({\rm dln}R/{\rm dln}M)_{\rm eq} = \zeta_{\rm L_{2}}$. Combined with the critical mass ratios \citep{ge10,ge15,ge19}, we systemically obtain the possible evolutionary channel, dynamically stable/unstable mass transfer, thermally stable/unstable mass transfer, for a given binary system with known initial mass, initial mass ratio, and initial orbital period. 

We describe how we build and solve the thermal equilibrium mass loss model in section 2. In section 3, we first take two $3.2~M_\odot$ ($Z=0.02$) stellar models, which are a MS ($R=2.99~R_\odot$) model with a radiative envelope, and an Asymptotic Giant Branch (AGB; $R=136.07~R_\odot$) model with a convective envelope,  as examples to show how the radius responds to thermal equilibrium mass loss. We secondly make a comparison of a $3.2~M_\odot$ ($Z=0.02$) terminal main sequence (TMS; $R=4.73~R_\odot$) and a RGB ($R=46.33~R_\odot$), and the stellar radius responses, to different mass-loss rates. Section 4 lists two different thresholds for thermal timescale mass transfer and shows the possible evolutionary channels of a binary system with a $3.2~M_\odot$ (AGB; $R=206~R_\odot$) lobe filling donor star. Section 5 presents the possible evolutionary channels for binary systems containing a $3.2~M_\odot$ donor star in evolutionary stages from the MS to the AGB by combining the different critical mass ratios. Section 6 summarizes the threshold conditions for thermal timescale mass transfer with donor stars mass from $0.1$ to $100~M_\odot$. We give a discussion and a summary in the last two sections.

\section{Thermal equilibrium mass loss model}
\label{thermal}

We are trying to examine the opposite extreme from that posed in our adiabatic mass loss papers~\citep{ge10,ge15,ge19}. In other words, the thermal equilibrium mass loss model describes an asymptote in the limit of arbitrarily {\it slow} mass transfer, instead of arbitrarily rapid mass loss. Ideally, the thermal equilibrium mass loss models represent the responses of stars to mass loss in the limit that mass loss is slow enough that donor stars are not driven out of thermal equilibrium. In reality, stars are never in complete thermal equilibrium. In the interest then of continuity between an evolutionary stellar model and its mass loss sequence, we fix the time derivative of the specific entropy, ${\rm d}s(m)/{\rm d}t$, not at zero (which imposes complete thermal equilibrium), but at its initial profile. We could instead have fixed the thermal energy generation rate ($\epsilon_{\rm gr}$) profile, as did \citet{hjel89b}, but that has some undesirable effects. For example, in an efficient convection zone, specific entropy, $s(m)$, is nearly independent of mass, $m$, which implies that $\epsilon_{\rm gr}=-T{\rm d}s/{\rm d}t$ should be proportional to the temperature at the local mass coordinate. By fixing ${\rm d}s(m)/{\rm d}t$, a smooth transition from the stellar model's previous evolution to the beginning of mass loss can guarantee that even a star in a different evolutionary stage is merely in complete thermal equilibrium. Hence, a stellar model's deviation from complete equilibrium is maintained.

For a better understanding of fixing ${\rm d}s(m)/{\rm d}t$, we start from the energy conservation (luminosity) equation as follows,
\begin{equation}
\frac{\partial L}{\partial m}= \epsilon_{\rm nuc} - \epsilon_\nu +\epsilon_{\rm gr}.
\label{luminosity1}
\end{equation}
Here $L$ is the luminosity, $m$ is the mass, $\epsilon_{\rm nuc}$ is the rate at which nuclear energy is produced per unit mass per second, $\epsilon_\nu$ is the rate at which the neutrinos take away energy per unit mass per second, and $\epsilon_{\rm gr}$ is the rate at which energy is absorbed ($\epsilon_{\rm gr} < 0$ typically in case of expansion) or released ($\epsilon_{\rm gr} > 0$ typically in case of contraction) per unit mass per second by the mass shell. Applying the combined first and second laws of thermodynamics, we get the expression of,
\begin{equation}
\epsilon_{\rm gr} =-\frac{\partial u}{\partial t} + \frac{P}{ \rho^2}\frac{\partial \rho}{\partial t}= -T \frac{\partial s}{\partial t}.
\label{grav-therm}
\end{equation}
In this formula, $u$ is the specific internal energy, $t$ is the time, $P$ is the pressure, $\rho$ is the density, $T$ is the temperature, and $s$ is the specific entropy. This $\epsilon_{\rm gr}$ describes changes to the thermal structures of stars, which change on the thermal timescales, $\tau_{\rm KH}$. If a star evolves on a much longer timescale than $\tau_{\rm KH}$, then $\epsilon_{\rm gr}$ approaches zero, and the star is in thermal equilibrium. To avoid the undesirable effects in efficient convection zones, we use ${\rm d}s(m)/{\rm d}t = -\epsilon_{\rm gr}/T$ as the parameter to descibe the thermal structure of a star. Hence, we fix ${\rm d}s(m)/{\rm d}t$ in the thermal equilibrium mass loss model.

For a star in both hydrostatic equilibrium and thermal equilibrium, the stellar structure equations become a set of ordinary differential equations. The normal stellar structure and evolution codes solve the simplified four first-order (radius, pressure, luminosity, and temperature) difference equations and a set of second-order (composition changes) difference equations. For a star in hydrostatic equilibrium, but not in thermal equilibrium, the time derivative represented by $\epsilon_{\rm gr}$ remains in the structure equations \citep[chap.\ 12.3]{kipp90}. We would have to specify the initial specific entropy profile $s_{\rm i}(m) = s(m,t_{\rm i})$ and know how the specific entropy changes with time $\dot{s}(m) = {\rm d} s(m)/{\rm d} t$. Hence, we build the adiabatic mass loss model \citep{ge10}, in which we assume that the specific entropy profile is fixed as the rapid mass transfer begins, to study the structure of stars undergoing very rapid mass transfer. On the contrary, if the mass-loss rate is comparable to or lower than its mass divided by its thermal timescale, $\dot{M} \lesssim M/\tau_{\rm KH}$, which means thermal relaxation is allowed, we assume the time derivative of the specific entropy profile $\dot{s}(m)$, which is related to Eq.~(\ref{grav-therm}), is fixed. With known $\dot{s}(m)$, the luminosity equation can become a first-order ordinary differential equation.

In summary, the thermal equilibrium mass loss model builds with two approximations. First, if a mass transfer process is on a comparable or faster timescale than the thermal timescale, we assume that the thermal relaxation response of the donor star is everywhere limited by the initial time derivative of the specific entropy profile, $\dot{s}_{\rm i}(m)$. Secondly, we assume that composition is frozen as a function of mass. Freezing ${\rm d}s/{\rm d}t$ as a function of mass means that these models are not strictly in thermal equilibrium (because stellar evolutionary models are never in complete thermal equilibrium), but it reproduces the gravo-thermal contribution to the stellar luminosity of the initial model of the mass-loss sequence, ensuring continuity of the mass-loss sequence. Freezing the composition proved necessary to prevent numerical diffusion from feeding fuel into regions previously exhausted but hot and dense enough to ignite any fuel so introduced. The stellar structure equations for the thermal equilibrium mass loss model have the same radius, pressure, and temperature equations, as the classical stellar structure equations, but with modified luminosity and composition change equations, as discussed above. 

\subsection{Stellar Structure Equations and Boundary Conditions}

We build the structure equations following the common practice in close binary evolutionary models, where we neglect the rotational and tidal effects on the structure of the donor star in a 1D spherically symmetric coordinate. Following the discussion above, we list the stellar structure equations of the thermal equilibrium mass loss model as follows. The four first-order ordinary differential equations of stellar structure, where the four structure variables are pressure $P$, radius $r$, temperature $T$, and luminosity $L$, vary with
an independent variable, mass $m$, are
\begin{equation}
\frac{d~{\rm ln}~P}{d~m}=-\frac{{\rm G}m}{4\pi r^4 P,}
\end{equation}
\begin{equation}
\frac{d~{\rm ln}~r}{d~m}=\frac{1}{4\pi r^3 \rho,}
\end{equation}
\begin{equation}
\frac{d~{\rm ln}T}{d~m}=\frac{d~{\rm ln}~P}{d~m} \nabla,\\
{\rm with} \quad \nabla  =  \left\{
\begin{array}{ll}\nabla_{\rm {rad}} \equiv \frac{3 \kappa P L}{16 \pi a c G m T^4} & \rm {Radiative\ zone}, \\
\nabla_{\rm {con}}   & \rm {Convective\ zone},
\end{array} \right.
\label{temperature}
\end{equation}
and 
\begin{equation}
\frac{d L}{d m}= \epsilon - \epsilon_\nu - T\frac{d s(m)}{d t},\\
{\rm with} \quad \frac{d s(m)}{d t} = \dot{s}(m)_{\rm i}.
\label{luminosity}
\end{equation}
Here density $\rho$, opacity $\kappa$, nuclear energy generation rate $\epsilon$, and neutrino loss rate $\epsilon_{\rm{\nu}}$, defined as usual, are functions of $P$, $T$, and the abundances $X_{\rm {n}}$ of various nuclear species. $G$, $c$ and $a$ are Newton's gravitational constant, the speed of light and the radiation constant, respectively. The subscript ${\rm i}$ in the time derivative of entropy means to keep the initial profile when the mass transfer begins.

Four boundary conditions are required to close this set of equations. At the center ($m=0$), 
\begin{equation}
r=0,
\end{equation}
\begin{equation}
L=0;
\end{equation}
and at the surface ($m=M$),
\begin{equation}
L=\pi a c r^2  T^4 ,
\end{equation}
\begin{equation}
\frac{\kappa \left(P_{\rm gas} + \frac{1}{2}P_{\rm rad}\right) }{g}=\frac{2}{3},
\end{equation}
where the latter surface boundary condition approximates a
classical gray atmosphere. The four equations are solved for
a given distribution of $X^{\rm i}_{\rm n} (m)$, which are fixed during mass loss process as we mentioned above.

\subsection{Numerical Implementation}

The numerical code to solve the equations of the thermal equilibrium mass loss model is written in $FORTRAN95$, based on the stellar evolution code developed by \citet{eggl71,eggl72,eggl73}, and \citet{paxt04}. The input physics, the equation of state \citep{eggl73b,vard60,webb75,pols95}, nuclear reaction rates and neutrino energy loss rates \citep{caug85,caug88,grab73,itoh96}, opacity tables \citep{alex94a,alex94b,itoh83,roge92}, etc., are also described by \citet{han94,han03} and \citet{pols95,pols98}. The essential features of the stellar evolution code have been kept in our thermal equilibrium mass loss model. These are (1) the use of an adaptive, moving, non-Lagrangian mesh, (2) the treatment of both convective and semi-convective mixing as diffusion processes and (3) the simultaneous, implicit solution of both the stellar structure equations and (in the evolutionary models) the chemical composition equations, including convective mixing. This code for the thermal equilibrium mass loss model is a parallel code to the adiabatic mass loss model in papers by \citet{ge10,ge15,ge19}. A detailed description of the numerical methods and input physics is given in those papers.

The use of an adaptive, moving, non-Lagrangian mesh naturally introduces two more parameters, the mesh point number, $k$, and the mesh function, $Q$. Mesh point number $k$ is distributed at uniform intervals in the function, $Q$, of the local structure variables \citep{eggl71}. The motivation for introducing $Q$ is that structure variables distribute monotonically and smoothly from the center to the surface of the stellar model with finite mesh point numbers. In other words, the important thing of introducing $Q$ is to control structure variables where things change most rapidly. Its introduction involves rewriting the stellar structure equations in terms of $k$, the mesh point number, as the independent (radial) variable, and introducing an additional structure equation for the mesh function,
\begin{equation}
\frac{{\rm d} Q}{{\rm d} k}= \frac{Q(N)-Q(1)}{N-1},
\end{equation}
where $N$ is the number of mesh points in the model. In our case,
we employ $800$ or $1000$ mesh points, distributed according to the mesh
function $Q$, which is the same as that used by \citet{ge10}.

In this code, we use ${\rm ln} f$ (a degeneracy parameter related to the
electron chemical potential; see \citet{eggl73,eggl06}, ${\rm ln} (L/(10^{33} {\rm erg~s^{-1}}))$ and ${\rm ln}(T/{\rm K})$ as our state variables, and ${\rm ln}(r/(10^{11} {\rm cm}))$, $m/(10^{33} {\rm g})$, and $Q$ as our global variables. The six differential equations for $P(k)$, $r(k)$, $m(k)$, $T(k)$, $L(k)$, and $Q(k)$ are written in different forms to usual. These are $PQ=0.05{\rm ln}(P/({\rm dyn~cm^{-2}}))+0.15{\rm ln}[(P/({\rm dyn~cm^{-2}}))+10^{15})]$, $RQ=-0.05 {\rm ln}(1+(r/(10^9{\rm cm}))^2)$, $MQ={\rm ln}[0.02m_c^{2/3}/(0.02m_c^{2/3}+m^{2/3})]$ with $m_c=3.5\rho(P/G/\rho^2)^{3/2}$, $TQ=0.45{\rm ln}[T/(T+2\times 10^4{\rm K})]$, $LQ=L\times(0.02m_c^{2/3}+m^{2/3})^{1/2}/m^{1/3}$, and $Q=PQ+RQ+MQ+TQ$, respectively. These new forms are designed to have the correct limiting behavior as $r$ approaches 0 and to avoid numerical problems at the surface. This new set of differential equations needs updated boundary conditions. At the surface ($k=1$), 
\begin{equation}
L=\pi a c r^2  T^4 ,
\end{equation}
\begin{equation}
\frac{\kappa \left(P_{\rm gas} + \frac{1}{2}P_{\rm rad}\right) }{g}=\frac{2}{3},
\end{equation}
\begin{equation}
m=M.
\end{equation}
Mass can be lost by modifying the surface mass boundary condition.
At the center ($k=N$), to avoid singular behavior as parameters approach 0, the central mesh point is offset so that
\begin{equation}
m=-\frac{{\rm d}m}{{\rm d}k},
\end{equation}
\begin{equation}
r=-\frac{{\rm d}r}{{\rm d}k},
\end{equation}
and
\begin{equation}
L=-\frac{{\rm d}L}{{\rm d}k}.
\end{equation}
The new composition equations are solved by fixing the composition profile. The differential structure equations are replaced by difference approximations with the same form as written by \citet{eggl71}. A Jacobian matrix is built for all the variables at different mesh points. The equations are then solved with {\it LAPACK} (www.netlib.org/lapack) and Newton-Raphson method.

\section{Response of $3.2 M_{\odot}$ Stars to thermal equilibrium mass loss}

\begin{figure}[ht!]
	\centering
	\includegraphics[scale=0.6]{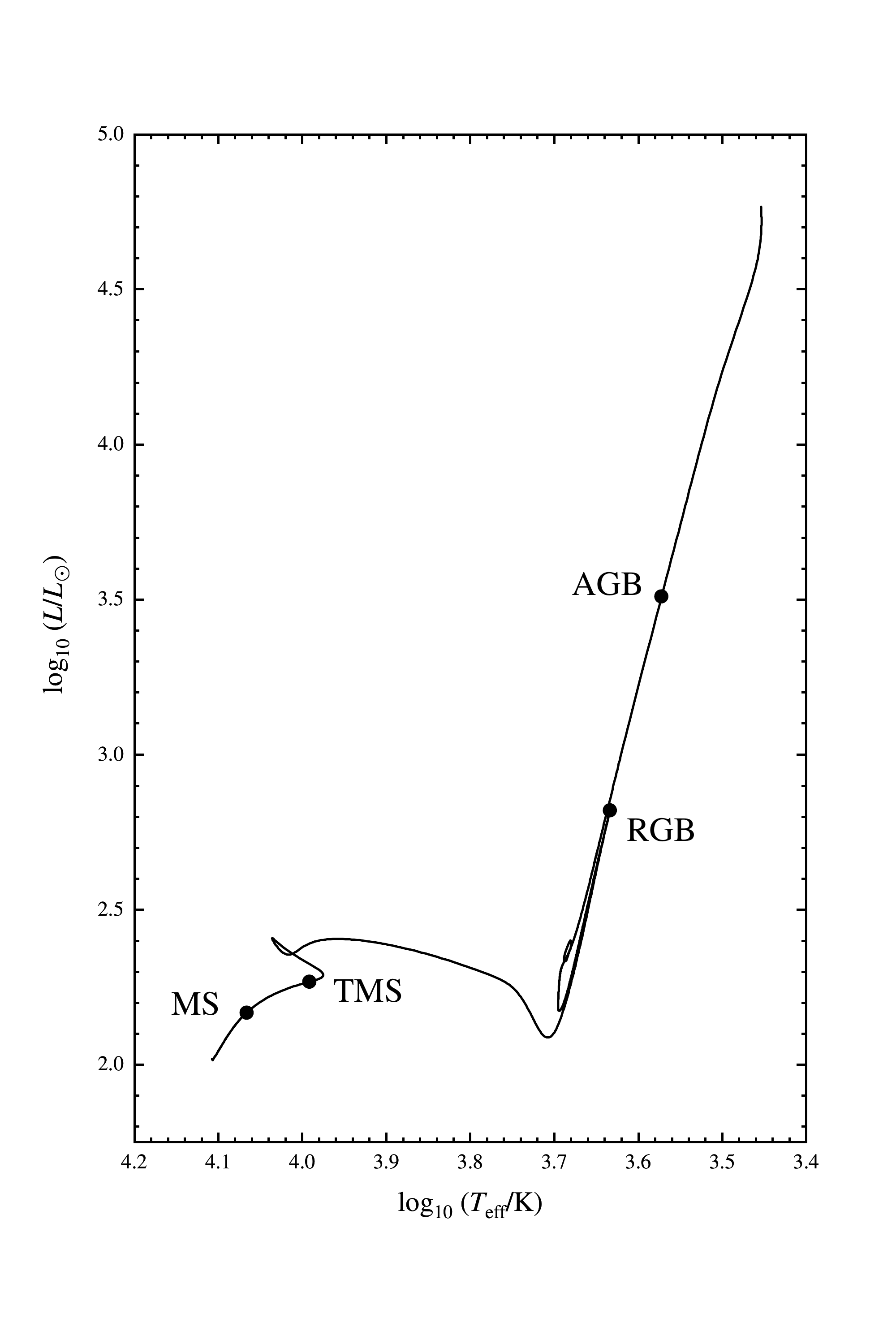}
	\caption{Examples of four stellar models on a $3.2~ M_{\odot}$ evolutionary track in a Hertzsprung-Russel diagram. \label{fig00}}
\end{figure}

\begin{figure}[ht!]
	\centering
	\includegraphics[scale=0.5]{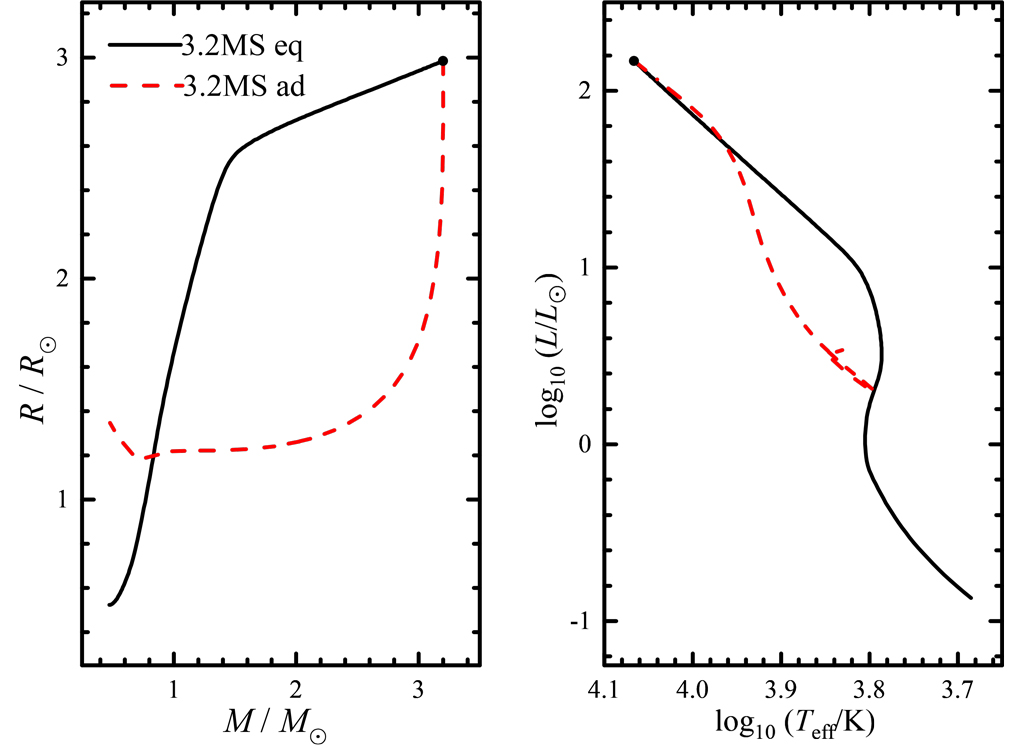}
	\caption{The response of a $3.2~ M_{\odot}$ and $2.99~R_{\odot}$ main-sequence (MS) star to thermal equilibrium (eq) mass loss (solid black line) and the adiabatic (ad) mass loss (red dashed line). The surface radius as a function of the remnant mass and the donor star's evolution in the Hurtzsprune-Russel diagram are shown in the left and the right panels, respectively. Black dot marks the start point of mass loss in each panel.
 \label{fig01}}
\end{figure}

\begin{figure}[ht!]
	\centering
	\includegraphics[scale=0.5]{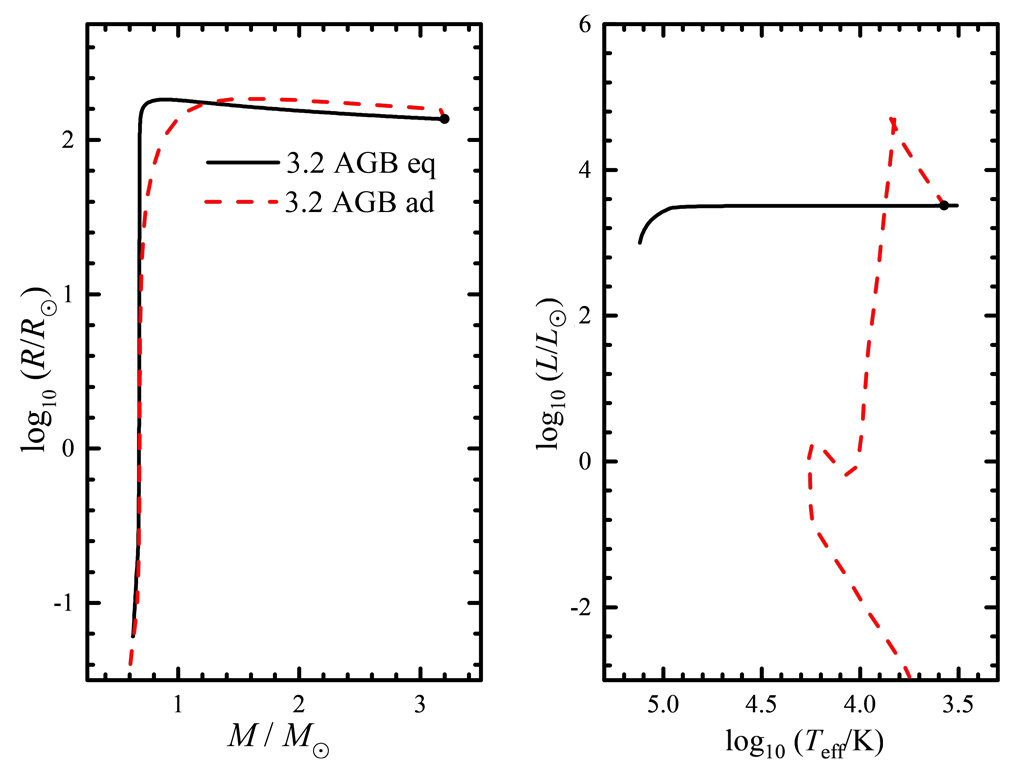}
	\caption{Similar to Figure~\ref{fig01}, but for the $3.2 M_{\odot}$ donor at Asymptotic Giant Branch (AGB, $R=136.07~R_{\odot}$). 
\label{fig02}}
\end{figure}

We first introduce how some of the global parameters of $3.2~M_{\odot}$ ($Z=0.02$) stellar models ($2.99~R_{\odot}$ MS and  $136.07~R_{\odot}$ AGB) respond to thermal equilibrium mass loss, and the differences to adiabatic mass loss. We secondly give the different mass-loss prescriptions for a $3.2~M_{\odot}$ TMS ($R=4.73~R_{\odot}$) and a $3.2~M_{\odot}$ RGB ($R=46.33~R_{\odot}$) star. To provide some context regarding the evolutionary stages of these initial stellar models, we show where the MS, TMS, RGB, AGB models lie on a $3.2~ M_{\odot}$ evolutionary track in a Hertzsprung-Russel diagram (Figure~\ref{fig00}).

If we assume that the response of a donor star to mass transfer is on an intermediate timescale, the entropy changing would be allowed within the star. The intermediate timescale mass transfer problem in binaries can now be treated in terms of thermal equilibrium mass loss from a single star. As we described in the last section, the approach is trying to maintain an initial stellar model's state of {\it disequilibrium}, in which the time derivative of the specific entropy profile $\dot{s}(m)$ is constant, as mass is removed. 

The cores and envelopes of stars respond in different ways between the thermal equilibrium mass loss and the adiabatic mass loss. Unlike the adiabatic mass loss model, in which heat flow is forbidden in rapid enough mass transfer, thermal relaxation is allowed in the thermal equilibrium mass loss model. So the changes in temperature and luminosity affect the hydrostatic equilibrium, the pressure and radius. The relative changes of the radius of the remnant star in thermal equilibrium mass loss are not as dramatic as in adiabatic mass loss (see the left panels in Figure \ref{fig01} and Figure \ref{fig02}).

Let us first check the stellar radius response to thermal equilibrium mass loss (the left panels in Figures~\ref{fig01} and~\ref{fig02}). In Figure~\ref{fig01}, this $3.2 M_{\odot}$ ($R=2.99~R_{\odot}$) MS star has a radiative envelope. So the mass layers expand more (the slope of the solid black line is shallower than that of the red dashed line) as the entropy increases because of the thermal relaxation ($\dot{s}$ profile is nonzero) and shrink less than for adiabatic mass loss compared with the initial radius. When the initial convection core ($M^{\rm i}_{\rm con}=0.50~ M_{\odot}$) of this $3.2~M_{\odot}$ MS star is nearly exposed, the radius responds differently in two different mass loss models. The radius continuously decreases in the thermal equilibrium mass loss model but starts to increase in the adiabatic mass loss model after the remnant mass becomes less than $0.75 M_{\odot}$. The radius responds differently after the remnant mass is less than $0.75 M_{\odot}$ in two models. The different response is because the convective core is allowed to shrink continuously in the whole thermal equilibrium mass loss process, and the convective core is located at the same mass in the adiabatic mass loss process. From the left panel in Figure~\ref{fig02}, we notice that in in the case of a $3.2~M_{\odot}$ ($R=136.07~R_{\odot}$) AGB star with a convective envelope, the mass layer initially expands less than in the case of adiabatic mass loss. The different radius response is because that the superadiabatic region is re-balanced when the surface move inwards in mass. However, after $2.0~M_{\odot}$ of the envelope is lost in the thermal equilibrium mass loss model, the heat flow from the inner part of the donor helps the remnant mass to expand more than in the adiabatic mass loss model. The mass at the bottom of the convective envelope, $M^{\rm i}_{\rm con}=0.684~ M_{\odot}$, is unchanged in adiabatic mass loss process, and it vanishes if the whole convective envelope is lost. However, the mass at the bottom of the convection envelope, $M_{\rm con}$, is allowed to change slightly in the thermal equilibrium mass loss process, and $M_{\rm con}$ finally falls sharply once the remnant mass is less than about $1.0~M_{\odot}$. 

We secondly check the luminosity and temperature response to thermal equilibrium mass loss (right panels in Figures~\ref{fig01} and~\ref{fig02}). In Figure~\ref{fig01}, the radiative envelope mass layers of the $3.2~ M_{\odot}$ MS star expand, although the surface radius shrinks compared to the star's initial radius. So the densities and temperatures decrease at mass layers throughout the radiative envelope. Thermal relaxation allows the core to contract to balance the core decompression and maintain the support of the envelope. The temperature gradient at newly exposed layers remains nearly equal to those in the initial mass layers. Hence, in the right panel of Figure~\ref{fig01}, we see the luminosity decreases gradually as the temperature decreases. The radius response to thermal equilibrium mass loss differs between the radiative envelope and the convective envelope. In Figure~\ref{fig02}, as the convective envelope mass layers of the $3.2~M_{\odot}$ AGB star expand, the density and temperature in principle should decrease initially. However, the convection transports the luminosity efficiently enough in the thermal equilibrium mass loss model after the tiny surface layer is lost. So we see a constant luminosity, slightly increasing radius, and dramatically increasing temperature in the right panel of Figure~\ref{fig02}. The luminosity finally starts to decrease after the remnant mass approaches the inner burning shell.

\begin{figure}[ht!]
	\centering
	\includegraphics[scale=0.64]{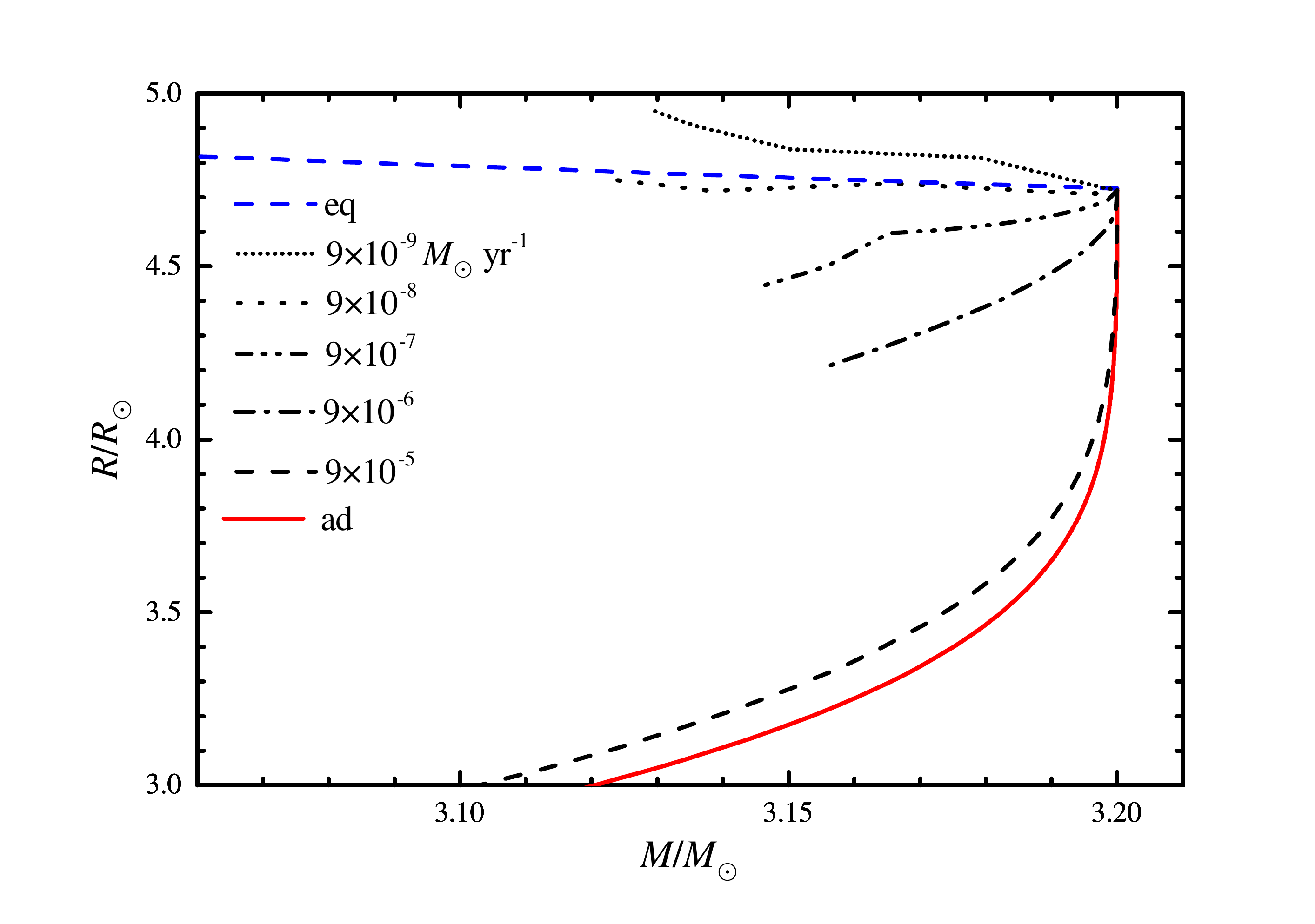}
	\caption{Radius response to different timescale, adiabatic (ad), and thermal equilibrium (eq) mass loss of a $3.2~M_{\odot}$ terminal main sequence (TMS) star. The red solid line and blue dashed line show the radius response to adiabatic mass loss and thermal equilibrium mass loss, respectively. Black lines represent how the radius responds to time-dependent calculations with constant mass loss rates at $9\times 10^{-9}$, $9\times 10^{-8}$, ..., $9\times 10^{-5}$ $M_{\odot} {\rm yr^{-1}}$. For this $3.2~M_{\odot}$ and $4.73~R_{\odot}$ TMS star, the stellar radius response to thermal equilibrium mass loss lies between thermal timescale and nuclear timescale mass loss ($\dot{M}_{\rm nuc} \simeq -9\times 10^{-9}~M_{\odot} {\rm yr^{-1}}$; $\dot{M}_{\rm KH} \simeq -9\times 10^{-6}~M_{\odot}/{\rm yr^{-1}}$). We may notice that some time-dependent calculations are stopped due to numerical reasons. \label{fig03}}
\end{figure}

\begin{figure}[ht!]
	\centering
	\includegraphics[scale=0.64]{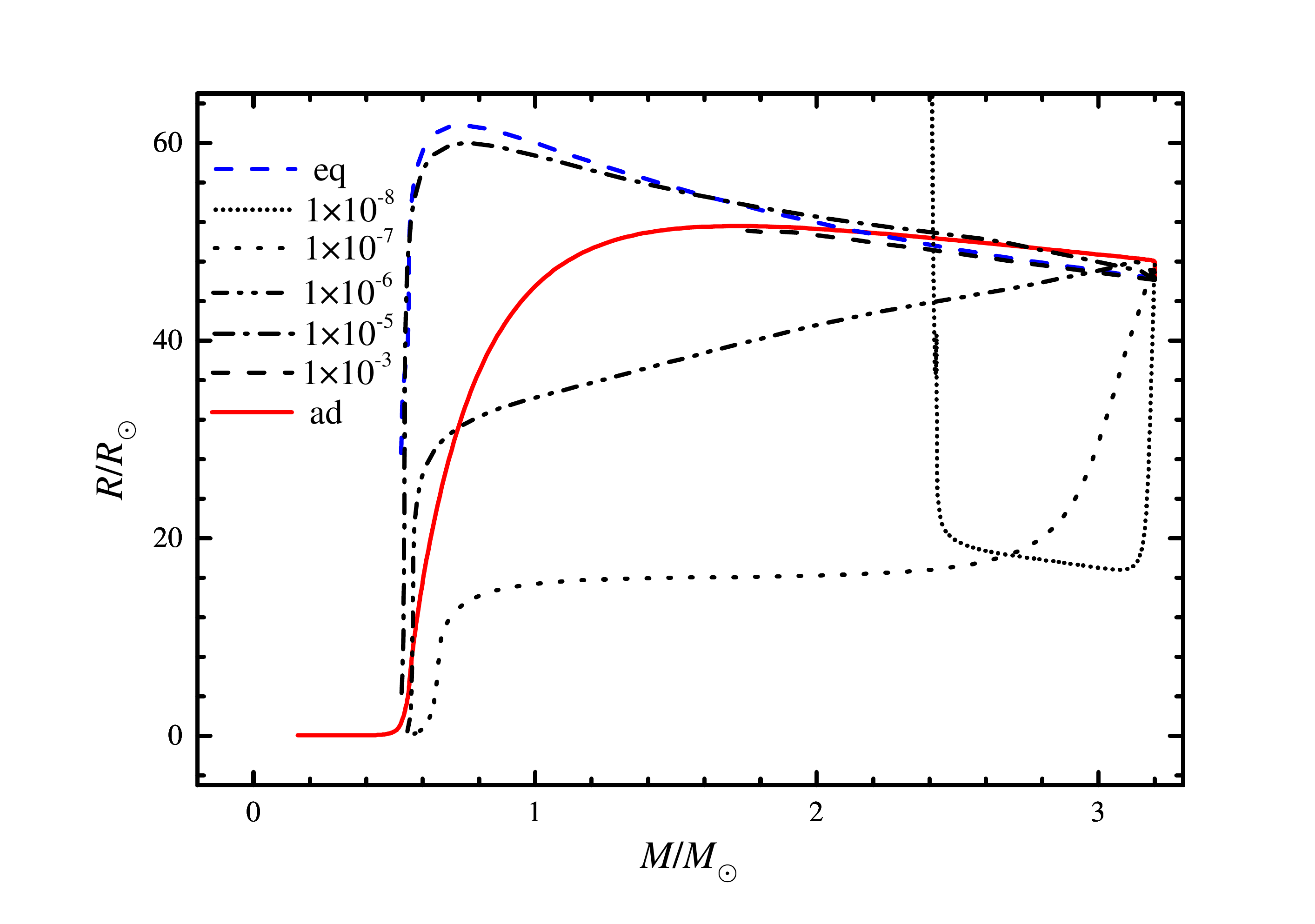}
	\caption{Radius response to different timescale, adiabatic (ad), and thermal equilibrium (eq) mass loss. The red solid line and blue dashed line have the same meaning as in Figure~\ref{fig03}. Black lines represent how the radius responds to time-dependent calculations with constant mass loss rates at $1\times 10^{-8}$, $1\times 10^{-7}$, ..., $1\times 10^{-3}$ $M_{\odot}{\rm yr^{-1}}$. For this $3.2~M_{\odot}$ and $46.33~R_{\odot}$ red giant branch (RGB) star, the stellar radius response to thermal equilibrium mass loss is much closer to thermal timescale mass loss ($\dot{M}_{\rm KH} \simeq -3\times 10^{-4}~M_{\odot}{\rm yr^{-1}}$) instead of nuclear timescale mass loss ($\dot{M}_{\rm nuc} \simeq -3\times 10^{-8}~M_{\odot}{\rm yr^{-1}}$). In the time-dependent calculation, we find that the dotted line evolves off the region because the mass-loss rate is too low, and the dashed line is stopped due to a numerical failure because the mass loss rate is too high.\label{fig04}}
\end{figure}

With the general thermal responses of the donor star in different structures (radiative layers or convective layers) in mind, we make a comparison of between a $3.2~M_{\odot}$ TMS and RGB stellar models' radii responses to mass loss on different mass loss rates from time-dependent calculations, besides our thermal equilibrium and adiabatic mass loss calculations (see both Figures~\ref{fig03} and \ref{fig04}). As for this $3.2~M_{\odot}$ TMS star with a radiative envelope, the radius is equal to $4.73~R_{\odot}$. Because MS stars are nearly in complete thermal equilibrium, $\dot{s}$ and thermal luminosity $L_{\rm th} = \int_0^M \epsilon_{\rm gr} {\rm d}m$ are both very small. From Figure~\ref{fig03}, we can expect that the stellar radius response to thermal equilibrium mass loss is between that for thermal timescale and nuclear timescale mass loss, but much closer to nuclear timescale mass loss ($\dot{M}_{\rm nuc} = -M_{\rm i}/\tau^{\rm i}_{\rm nuc} \simeq -9\times 10^{-9}~M_{\odot}{\rm yr^{-1}}$) instead of thermal timescale mass loss ($\dot{M}_{\rm KH} = -M_{\rm i}/\tau^{\rm i}_{\rm KH} \simeq -9\times 10^{-6}~M_{\odot}{\rm yr^{-1}}$). We also take a $3.2~M_{\odot}$ RGB star with a convective envelope, which has a $46.33~R_{\odot}$ radius, as an example. Unlike MS stars, RGB stars are evolving very fast along the giant branch. So $\dot{s}$ and thermal luminosity $L_{\rm th}$ cannot be neglected. The stellar radius response to thermal equilibrium mass loss is between that of thermal timescale and nuclear timescale mass loss, but much closer to thermal timescale mass loss ($\dot{M}_{\rm KH} = -M_{\rm i}/\tau^{\rm i}_{\rm KH} \simeq -3\times 10^{-4}~M_{\odot}{\rm yr^{-1}}$) instead of nuclear timescale mass loss ($\dot{M}_{\rm nuc} = -M_{\rm i}/\tau^{\rm i}_{\rm nuc} \simeq -3\times 10^{-8}~M_{\odot}{\rm yr^{-1}}$). From the comparison above, it is safe to conclude that the adiabatic mass loss model gives a good approximation to donor stars suffering faster than thermal timescale mass transfer, and that the thermal equilibrium mass loss model shows the response of donor stars undergoing intermediate timescale mass loss between nuclear and thermal timescale mass loss. We introduce how we apply the radius response to thermal equilibrium mass loss to search for the thresholds for thermal timescale mass transfer in the next section.

\section{Thresholds for thermal timescale mass transfer}

We address here the application of thermal equilibrium mass loss sequences on evaluating the threshold conditions for the onset of thermal timescale (th) mass transfer. From the classical radius-mass exponent definition, $\zeta= {\rm dln} R/{\rm dln} M$, we can define the tide (Roche-lobe, L), thermal equilibrium (eq), adiabatic (ad) radius-mass exponent, respectively, as follows:
\begin{equation}
\zeta_{\rm L}=\frac{{\rm dln}R_{\rm L}}{{\rm dln}M}
\end{equation} 
\begin{equation}
\zeta_{\rm eq}=\frac{{\rm dln}R_{\rm eq}}{{\rm dln}M}
\end{equation} 
\begin{equation}
\zeta_{\rm ad}=\frac{{\rm dln}R_{\rm ad}}{{\rm dln}M}
\end{equation} 
Three different timescale mass transfer can be classified by comparing these three radius-mass exponents (Webbink 1985, Ge et al 2010), as $\zeta_{\rm ad}< \zeta_{\rm L}$ for dynamical timescale mass transfer, $\zeta_{\rm eq}< \zeta_{\rm L}< \zeta_{\rm ad}$ for thermal timescale mass transfer, and $\zeta_{\rm L}< (\zeta_{\rm eq},\zeta_{\rm ad})$ for nuclear timescale mass transfer. 

Starting with the simplest assumption, we assume mass transfer between binary components is fully conservative, by which we mean specifically that the total mass of the binary
\begin{equation}
M_{\rm tot} = M_1 + M_2
\label{mtot}
\end{equation}
and orbital angular momentum of the binary
\begin{equation}
J_{\rm orb} = \left( \frac{M_1^2 M_2^2}{M_1 + M_2}\,GA \right)^{1/2}
\label{jorb}
\end{equation}
are constant. Here, $M_1$ is the mass of the donor, $M_2$ is the mass of the accretor, $G$ is the gravitational constant, and $A$ is the separation between the two components. Then, we use Eggleton's~\citep{eggl83} approximation for the Roche-lobe radius,
\begin{equation}
r_{\rm L}(q) = \frac{0.49 q^{2/3}}{0.6 q^{2/3} + \ln(1+q^{1/3})} = \frac{R_{\rm L}}{A}\ ,
\label{rlobe}
\end{equation}
where the mass ratio $q \equiv M_1/M_2$ is defined to be the ratio of donor star mass to accretor mass. At the beginning of mass transfer, the donor star just fills in its Roche-lobe. So we have
\begin{equation}
R_{{\rm Li}}= {R_{\rm i}}.
\label{rlobe_rstar_i}
\end{equation}
Under conservative mass transfer, the mass ratio, $q$, when the donor star has been reduced from mass $M_{\rm i}$ to mass $M_1$, is
\begin{equation}
q =\left[\frac{M_{\rm i}}{M_1}(1+q_{\rm i}^{-1})-1\right]^{-1},
\end{equation} 
where $q_{\rm i}$ is again the initial mass ratio (donor/accretor). The binary separation at this point is 
\begin{equation}
A(q) = A_{\rm i}\left(\frac{1+q}{1+q_{\rm i}}\right) ^4\left(\frac{q_{\rm i}}{q}\right)^2 \ .
\label{seperation}
\end{equation} 
Combining Eqs. (\ref{rlobe})-(\ref{seperation}), we have
\begin{equation}
R_{\rm L}(q)= R_{{\rm Li}} \frac{r_{\rm L}(q)}{r_{\rm L}(q_{\rm i})}\left(\frac{1+q}{1+q_{\rm i}}\right) ^4\left(\frac{q_{\rm i}}{q}\right)^2 \ .
\label{rlobe_rstar}
\end{equation}
We can write the Roche-lobe radius-mass relation directly in terms of the binary mass ratio:
\begin{equation}
\zeta_{\rm L} \equiv \left(\frac{\partial \ln R_{\rm L}}{\partial \ln M_1}\right)_{J,M} = \left[\frac{2\ln(1+q^{1/3})-q^{1/3}/(1+q^{1/3})}{3[0.6q^{2/3} + \ln(1+q^{1/3})]}-2\left(\frac{1-q}{1+q}\right) \right](1+q)\ .
\label{zetalobe}
\end{equation}
Given $\zeta_{\rm eq}$ from the thermal equilibrium mass loss sequence for a donor star of interest, Eq.~(\ref{zetalobe}) then implicitly defines a corresponding critical mass ratio, $q_{\rm eq}$, satisfying the equation
\begin{equation}
\zeta_{\rm eq} = \zeta_{\rm L}(q_{\rm eq})
\label{qcrit_zeta}
\end{equation}
above which a binary containing that donor star is unstable to thermal timescale mass transfer.

However, we should notice that the radius-mass exponents, $\zeta_{\rm L}$, $\zeta_{\rm ad}$, and $\zeta_{\rm eq}$, are not constant throughout the mass transfer, but vary continuously, in some cases very rapidly. So we should track down the whole mass loss process to find the minimum initial mass ratio, $q_{\rm eq}$, such that $\zeta_{\rm eq}\le \zeta_{\rm L}$ always remains, as the critical limit for thermal timescale mass transfer. An alternative indication of thermal timescale (th) mass transfer is that the donor star overfills its inner Lagrangian radius deeply enough. So we do not just compare the surface radius, $R$, to its Roche-lobe radius, $R_{\rm L}$, to calculate the critical mass ratio $q_{\rm eq}$. Instead, we consider the Roche-lobe radius at its deepest penetration into the donor star at $R_{\rm KH}$. The desired threshold condition for thermal timescale (th) mass transfer then consists of finding the initial mass ratio, $q_{\rm th}$, which suffices to drive mass transfer at a thermal (Kelvin-Helmholtz) rate. In other words, we compare the inner radius $R_{\rm KH}$ to the Roche-lobe radius $R_{\rm L}$ during thermal equilibrium mass loss to find the critical mass ratio $q_{\rm th}$ from $\zeta_{\rm th} \equiv ({\rm dln}R_{\rm KH}/{\rm dln}M)_{\rm eq} = \zeta_{\rm L}$. It may arise that the donor overfills its Roche-lobe so deeply that it even overfills its outer Lagrangian point ${\rm L_{2}}$. In such a situation, the mass could be lost from (the outer Lagrangian points of) the system. But that material may still stay in the vicinity of the system as circumbinary material for a while. The binary system might enter a {\it contact like} or {\it common envelope like} phase. We compare the radius, $R$, to its Roche-lobe radius, $R_{\rm L_2}$, at outer Lagrangian point $L_2$ during thermal equilibrium mass loss to find the critical mass ratio $q_{\rm L_2}$ from $\zeta_{\rm eq} \equiv ({\rm dln}R/{\rm dln}M)_{\rm eq} = \zeta_{\rm L_{2}}$. We describe the corresponding methods in the next two subsections and apply these to a $3.2 M_{\odot}$ AGB star, shown on a radius vs. remnant mass diagram in the last subsection.

\subsection{Critical Mass Ratio $q_{\rm th}$}

In the thermal equilibrium sequences, matter streamlines through the ${\rm L_1}$ region do not behave adiabatically, as assumed in the adiabatic mass loss sequences \citep{ge10}, because thermal relaxation timescales across the flow might be shorter than the local dynamical timescale (as emphasized by~\citet{wood11} in case of convective stars). We therefore assume, as a matter of expediency, that the flow retains the same thermal structure (i.e., temperature) as a function of potential as it would have far from ${\rm L_1}$, at the origin of the stream (by analogy with the specific entropy profile preserved in the adiabatic mass loss sequences by \cite{ge10}. Hence, we assume that temperature, rather than specific entropy, remains constant along the isothermal streamlines. Then, in place of Eq. (A9) of \cite{ge10}, namely
\begin{equation}
\dot{M_1} =-\frac{2\pi R^3_{\rm L}}{G M_1} F(q)\int_{\phi_{\rm L}}^{\phi_s} \Gamma_1^{1/2}\left(\frac{2}{\Gamma_1 + 1}\right)^{\frac{\Gamma_1 + 1}{2(\Gamma_1 - 1)}}(\rho P)^{1/2}
{\rm d}\phi,
\end{equation}
where
$\Gamma_1 \equiv (\partial {\rm \ln}P/\partial {\rm \ln}\rho )_s$ is the first adiabatic exponent and $M_1$ is the donor star, we write the isothermal mass loss formula instead
\begin{equation}
\dot{M_1} =-\frac{2\pi R^3_{\rm L}}{G M_1} F(q)\int_{\phi_{\rm L}}^{\phi_s} \chi_\rho^{1/2}\left(\frac{2}{\chi_\rho + 1}\right)^{\frac{\chi_\rho + 1}{2(\chi_\rho - 1)}}(\rho P)^{1/2}
{\rm d}\phi,
\label{isothermal-ml}
\end{equation}
where
\begin{equation}
\chi_\rho \equiv \left(\frac{\partial {\rm \ln}P}{\partial {\rm \ln}\rho}\right)_{T,X}  
= \left(\frac{\partial {\rm \ln}P}{\partial {\rm \ln}f}\right)_{T,X}\times \left(\frac{\partial {\rm \ln}\rho}{\partial {\rm \ln}f}\right)^{-1}_{T,X}.
\end{equation}
Here, $f$ is the electron degeneracy parameter as a function of which we cast our equation of state. 

We define the inner radius, $R_{\rm KH}$, to be equal to the Roche-lobe radius, $R_{\rm L}$, in Eq.~(\ref{isothermal-ml}). We also define $R_{\rm i}$, $L_{\rm i}$, and $M_{\rm i}$, as the initial radius, luminosity, and mass, respectively, when mass loss begins. By solving $\dot{M_1}(R_{\rm KH}) = \dot{M}_{\rm KH}\equiv-R_{\rm i}L_{\rm i}/(GM_{\rm i})$, the thermal timescale (th)\footnote{To distinguish the difference between the thermal equilibrium mass loss and the thermal timescale mass loss, we list both the short abbreviations of eq and th in many places.} mass loss rate, along the thermal equilibrium (eq) mass loss process, we can calculate the $R_{\rm KH}$ as a function of its remnant mass. Combining with Eq.~(\ref{zetalobe}), we can finally find the minimum the critical mass ratio $q_{\rm th}$ from $\zeta_{\rm th} = ({\rm dln}R_{\rm KH}/{\rm dln}M_1)_{\rm eq} = \zeta_{\rm L}$.

\subsection{Critical Mass Ratio $q_{\rm L_2}$}

We take the critical mass ratio for the overflow of the outer Lagrangian surface, $q_{\rm L_2}$, to be that initial mass ratio such that the volume-equivalent radius of the outer lobe, $R_{\rm L_2}$, is tangentially equal to the stellar radius, $R$, at one point along a mass-loss sequence. 

Let $M_{\rm i}$ be the initial donor mass in a mass-loss sequence, $R_{\rm i}$ the initial donor radius, and $q_{\rm i} = M_{\rm i}/M_{\rm 2i}$ the initial mass ratio. $M_{\rm 2i}$ is the initial mass of the companion star. At the start of the mass-loss sequence, the donor just fills its Roche-lobe, so $R_{\rm i} = A_{\rm i} r_{\rm L}(q_{\rm i}) $, where $A_{\rm i}$ is the initial orbital separation. With $q = q_{\rm i}$, Eq.~(\ref{rlobe}) can give $r_{\rm L}(q)$. 

The dimensionless radius of the outer critical surface, $r_{\rm L_2}$, according to Webbink, can be approximated as 
\begin{equation}
r_{\rm L_2}(q) = r_{\rm L}(q) +\left[0.179 + 0.01\left(\frac{q}{1+q}\right)\right]\left(\frac{q}{1+q}\right)^{0.625}\  {\rm for}\ q \le 1\ ,
\label{rl2_1}
\end{equation}
or 
\begin{equation}
r_{\rm L_2}(q) = r_{\rm L}(q) +\left[0.179 + 0.01\left(\frac{q}{1+q}\right)-0.025\left(\frac{q-1}{q}\right)\right]\left(\frac{q}{1+q}\right)^{0.625} q^{-0.74} \  {\rm for}\ q \ge 1\ .
\label{rl2_2}
\end{equation} 
The physical radius of the outer Roche-lobe is then $R_{\rm L_2}(q) = A(q)r_{\rm L_2}(q) $, where $q$ is the current mass ratio in the mass-loss sequence, and $A(q)$ the current binary separation. Eqs. (\ref{rl2_1}) and (\ref{rl2_2}) are approximate analytic fits to integrations of the Roche limit by \citet{moch84} and \citet{penn85}. The relevant results are displayed in Tables 6 \& 7 of \citet{moch84} (we use the entries for F=2.0), and Appendix (Table A5) in \citet{penn85}. These papers characterize the outer critical surface in terms of their volume-equivalent radii (as does Eggleton's formula for the inner critical surface). Volume-equivalent radii are at best only rough approximations, as are the Roche models generally.

The above equations, including Eqs. (\ref{rlobe})-(\ref{seperation}), then form a complete, analytic relation for the calculation of $R_{\rm L_2}$, given the current mass $M_1$ of the donor star and its initial mass $M_{\rm i}$, mass ratio $q_{\rm i}$, and radius $R_{\rm i}$. Finding $q_{\rm L_2}$  then involves inverting this relation to find $q_{\rm i}$, given $M_{\rm i}$, $R_{\rm i}$, $M_1$, and $R = R_{\rm L_2}$ ($R$ being the donor radius at mass $M_1$ along the mass-loss sequence). The critical mass ratio $q_{\rm L_2}$ is the minimum $q_{\rm i}$ along the mass-loss sequence. 

\subsection{Critical Mass Ratios of a $3.2 M_{\odot}$ AGB Star}
\begin{figure}[ht!]
	\centering
	\includegraphics[scale=0.64	]{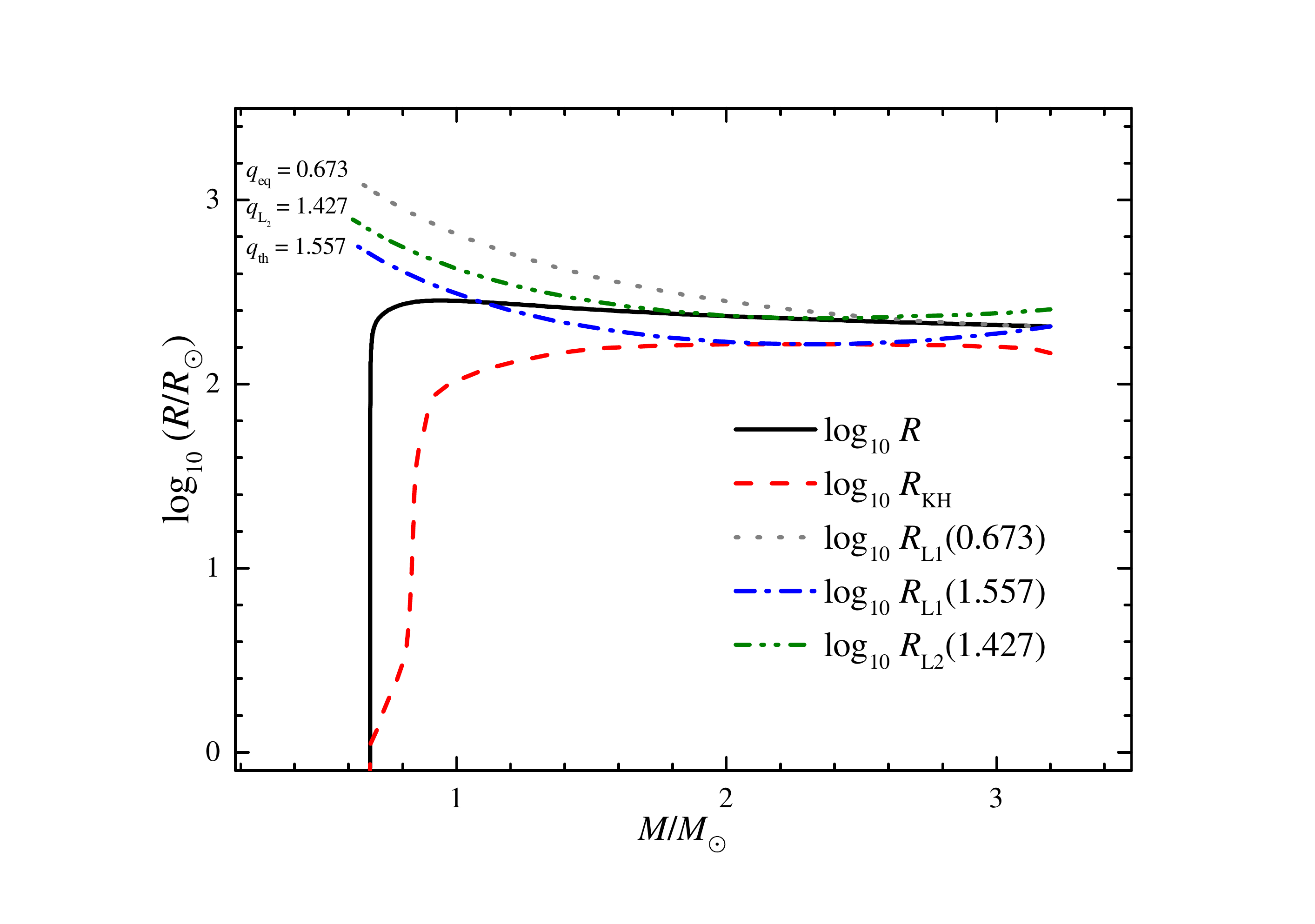}
	\caption{Different thresholds for thermal-timescale mass transfer applied to a $3.2 M_{\odot}$ AGB star ($R_{\rm i}=206~R_{\odot}$). Black solid line is the radius response to thermal-equilibrium mass loss; red dashed line shows the inner radius $R_{\rm KH}$ as a function of its remnant mass; blue dash dotted line gives the inner Roche-lobe radius $R_{\rm L_1}$ with an initial mass ratio $q_{\rm i}=1.557$; green dash dot-dotted line shows the outer Roche-lobe radius $R_{\rm L_2}$ with an initial mass ratio $q_{\rm i}=1.427$; and gray dotted line is the inner Roche-lobe radius $R_{\rm L_1}$ with an initial mass ratio $q_{\rm i}=0.673$. \label{fig05}}
\end{figure}

We take a $3.2 M_{\odot}$ AGB stellar model with an initial radius, $R_{\rm i}=206~ R_{\odot}$, as an example to show how these different critical mass ratios are competing. Figure \ref{fig05} shows how the surface radius of the donor, as well as the Roche lobe radius (conservative mass transfer) with given initial mass ratio, responds to thermal equilibrium mass loss. This binary system contains a $3.2 M_{\odot}$ AGB donor star that just over-fills its Roche-lobe with $R_{\rm i} = R_{\rm L_1} = 206 R_{\odot} $. If the initial mass ratio, $q_{\rm i}$, is larger than 1.557, thermal timescale mass transfer ensures because the critical initial mass ratio, $q_{\rm th} = 1.557$, is that at which the radius $R_{\rm KH}$ (red dashed line) is tangential to the Roche-lobe radius $R_{\rm L_1}$ (blue dash-dotted line). If the initial mass ratio, $q_{\rm i}$, is larger than 1.427, the donor star would overfill its outer Lagrangian point $L_2$. When $q_{\rm i}$ is equal to 1.427, the surface radius $R$ (solid black line) is tangent to its outer Roche-lobe radius $R_{\rm L_2}$ (green dash-dot-dot line). In this case, a common envelope might also form through the mass loss via the outer Lagrangian point. Besides $q_{\rm L_2}=1.427$, we also find the surface radius $R$ (solid black line) is tangent to its Roche-lobe radius $R_{\rm L_1}$ (gray dotted line) when $q = q_{\rm eq}=0.673$. Therefore, if the initial mass ratio, $q_{\rm i}$, is smaller than 1.427 and larger than 0.673, the mass transfer could be driven by either angular momentum loss or radius expansion caused by nuclear burning. If the initial mass ratio, $q_{\rm i}$, is smaller than 0.673 and there is no significant angular momentum loss, only nuclear timescale mass transfer is possible. With these different critical mass ratios, in principle, for a binary system that starts Roche-lobe overflow with known initial donor mass and mass ratio, we can know its likely mass transfer channel. We give a more detailed study in the next section for binary systems containing different $3.2~M_\odot$ donor stars.

\section{Mass transfer channels of $3.2~M_\odot$ donor stars}
\label{sec_mt3_2}

\begin{figure}[ht!]
	\centering
	\includegraphics[scale=0.36]{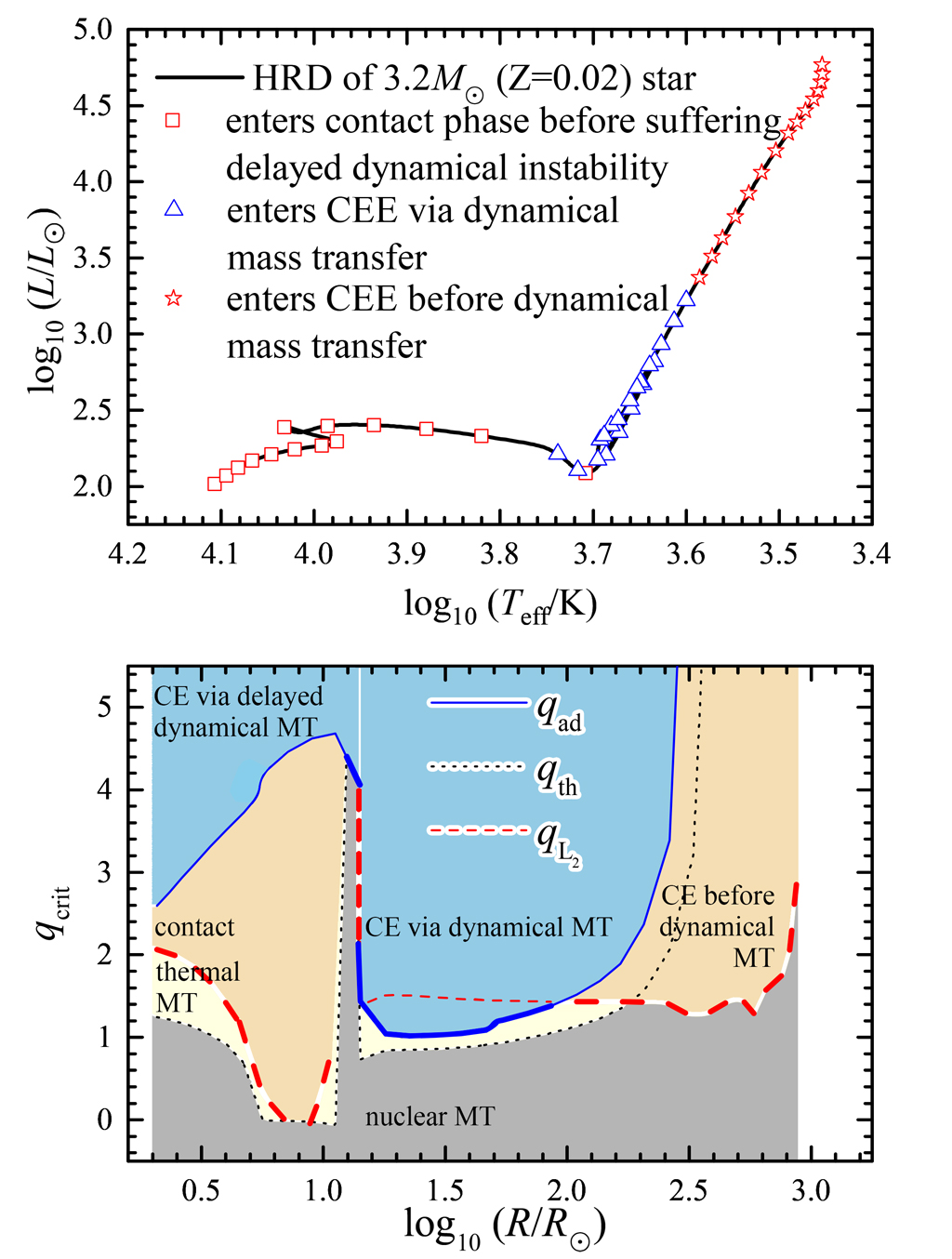}
	\caption{Possible evolution channels shown on the critical mass ratio vs. the initial stellar radius diagram (lower panel) and the Hertzsprung-Russel diagram (upper panel) for binary systems containing a $3.2 M_{\odot}$ donor star. The lower panel shows how the critical initial mass ratios change with the donor star's radius. The radius is increasing with evolution and that the colored regions correspond to the onset of mass transfer at that radius for the given mass ratio. Three different critical mass ratios, $q_{\rm ad}$, $q_{\rm th}$, and $q_{\rm L_2}$, are used to identify the boundaries between possible evolutionary channels, dynamical timescale mass transfer, thermal timescale mass transfer, and overfilling the outer Lagrangian surface ${\rm L_2}$. In the lower panel, different filling colors indicate what kind of mass transfer channel the donor star may suffer from for the initial mass ratio in that range. The thick red dashed line and thick solid blue line show the more effective constraint leading to a common envelope phase. For the clarity, we omit the stellar models at evolutionary phases in which the donor star's radius is smaller than in a preceding phase, such as during core helium burning, or just beyond the terminal main sequence. The red squares in the upper panel indicate the possible evolutionary channels of the MS and HG donors. These correspond to the thick red dashed line on the left of the lower panel. The blue triangles show the possible evolutionary channels of the RGB and early AGB donors. These correspond to the thick solid blue line in the middle of the lower panel. The red stars in the upper panel reveal the possible evolutionary channels of the late AGB donors, and they correspond to the thick red dashed line on the right of the lower panel.  \label{fig06}}
\end{figure}

With the critical initial mass ratios, $q_{\rm th}$, and $q_{\rm L_2}$, for thermal timescale mass transfer described in the above section and the critical initial mass ratio, $q_{\rm ad}$, for dynamical timescale mass transfer studied in the paper of \citet{ge19}, we can study the possible evolutionary channels for a given binary system. A binary system containing a Roche-lobe-filling donor star, which starts transferring mass to its companion on a given evolutionary stage is shown for given initial mass $M_{\rm i}$ and radius $R_{\rm i}$. Here, as an example, we address the possible evolution channels, dynamical timescale mass transfer ($q_{\rm i} > q_{\rm ad}$), thermal timescale mass transfer ($q_{\rm i} > q_{\rm th}$) and overfilling of the outer Lagrangian surface to begin a CE like process ($q_{\rm i} > q_{\rm L_2}$), of a binary system with a $3.2 M_{\odot}$ ($Z=0.02$) donor star.

Donor stars with a mass of $3.2 M_{\odot}$ have a radiative envelope on the MS and HG. The radiative envelope expands for more evolved donor stars. So the critical mass ratio for delayed dynamical timescale mass transfer, $q_{\rm ad}$, is large and increases with evolutionary stages (solid blue line in the lower panel in Figure \ref{fig06}). The envelope of the $3.2 M_{\odot}$ star expands even faster when it passes through the HG, and it only shrinks at the very last stage of the HG. The envelope of the donor star could keep its evolutionary expanding or shrinking state in a thermal equilibrium mass loss process. For these reasons, the critical mass ratio for thermal timescale mass transfer, $q_{\rm th}$ (dotted black line), decreases from the zero-age main sequence (ZAMS) to the late stage of the HG and increases dramatically at the very last HG stage in the lower panel of Figure~\ref{fig06}. For the same reasons, the critical mass ratio for overflowing the outer Lagrangian surface, $q_{\rm L_2}$ (red dashed line), shows the same pattern.

The $3.2 M_{\odot}$ donor stars on the RGB and AGB have a convective envelope. The convective envelope penetrates deeper into the RGB/AGB star as the star evolves. The donor stars on the RGB and early AGB have extended convective envelopes, and we might expect that the critical mass ratio for dynamical mass transfer is less than 1 \citep{hjel87}. However, the existence of non-ideal gas, partially ionized gas and inefficient convection in the donor star increase the critical mass ratio for dynamical timescale mass transfer, $q_{\rm ad}$, from around 1.0 to 1.5 (solid blue line in the lower panel of Figure \ref{fig06}, with ${\rm log} (R/R_\odot)$ from 1.2 to 2.2). The envelope becomes more extended for more evolved AGB stars; so the thermal timescale becomes shorter. The density and temperature in the envelope decrease as well for more evolved donors. So the donor star needs a larger and larger mass transfer rate to reach thermal timescale or dynamical timescale mass transfer. We, therefore, find both $q_{\rm ad}$ and $q_{\rm th}$ increase quickly, as shown in Figure~\ref{fig06} (solid blue line and dotted black line in the right part of the lower panel, respectively), after the donor star enters the thermally pulsing asymptotic giant branch (TPAGB) stage. Nevertheless, the critical mass ratio for overfilling its outer Lagrangian surface, $q_{\rm L_2}$, (red dashed line in the right part of the lower panel in Figure~\ref{fig06}) remains nearly constant for RGB stars and most AGB stars. Then, $q_{\rm L_2}$ increases from 1.4 to 2.93 for the TPAGB stars (see also in Table \ref{theqml}). We could argue that $q_{\rm L_2}$ becomes a more important and strict parameter for unstable mass transfer because $q_{\rm L_2}$ becomes smaller than both $q_{\rm ad}$ and $q_{\rm th}$ for ${\rm log(} R{\rm /}R_\odot{\rm )>} 2$. Hence, we show the dominant critical mass ratio (the smaller of $q_{\rm ad}$ and $q_{\rm L_2}$) for the common envelope phase with thick lines in the lower panel of Figure~\ref{fig06}. The corresponding results are also shown in the Hertzsprung-Russel diagram (the upper panel of Figure~\ref{fig06}) when the initial mass ratio is larger than the critical mass ratios.

Now based on Figure~\ref{fig06}, we could argue that $3.2 M_{\odot}$ MS and HG stars may enter a contact phase before suffering delayed dynamical instability ($q_{\rm L_2} < q_{\rm ad}$; red squares on the upper panel in Figure~\ref{fig06}), if the initial mass ratio, $q_{\rm i}$, is larger than $q_{\rm L_2}$ but smaller than $q_{\rm ad}$. The existence of the thick radiative envelope of the MS and HG donor stars makes the binary systems enter into a contact phase before the convective core of the donor is exposed. Meanwhile, these donor stars suffer a delayed dynamical instability if $q_{\rm i} > q_{\rm ad}$. The $3.2 M_{\odot}$ RGB and early AGB stars could enter a CE process via dynamical timescale mass transfer ($q_{\rm ad} < q_{\rm L_2}$; blue triangles in the upper panel of Figure~\ref{fig06}). Furthermore, $3.2 M_{\odot}$ late AGB (or TPAGB) stars may also enter a CE phase even before initially suffering dynamical timescale mass transfer ($q_{\rm L_2} < q_{\rm ad}$; red stars in the upper panel of Figure~\ref{fig06}) because even the thermal timescale mass transfer should be fast enough, not to mention the overfilling of the outer Lagrangian surface of the donor star, to cause the formation of a common envelope.

\section{Model Grid and Results}
\label{sec_res}

The initial stellar models for the thermal equilibrium mass loss sequences reported in this paper are constructed from a library of stellar evolution sequences with Population I metallicity ($Z=0.02$). The initial masses, spanning from $0.1$ to $100~M_{\odot}$, are constructed at intervals of $\Delta~{\rm log} M~\approx 0.1$ ($\Delta~{\rm log}~M \approx 0.05$ for low-mass stars). The initial radii are constructed at intervals of $\Delta~{\rm log}~R \approx 0.1$ from the donor stars on the MS, across the Hertzsprung gap, on the first giant branch, experiencing core helium burning, to the AGB stars and advanced stages of nuclear burning (if applicable). So the stellar model grid covers the full range of masses and evolutionary stages of potential interesting donor stars.

Tables~\ref{intmod} and~\ref{glbmod} document the initial properties of the donor stars at the beginning of each thermal equilibrium mass loss sequence. Table~\ref{intmod} is arranged in segments, by stellar mass, $M_{\rm i}$. Successive columns list

\begin{deluxetable}{rrrrrrrrrrr}   
	\tabletypesize{\footnotesize}
	\tablewidth{0pt}
	\tablecolumns{11}
	\tablecaption{Interior properties of initial models\label{intmod}}
	
	\tablehead{
		\colhead{$k$} & \colhead{$\log_{10} (\frac{t}{{\rm yr}})$} & \colhead{$\frac{M_{\rm ce}}{M_\sun}$} & \colhead{$\frac{M_{\rm c}}{M_\sun}$} & \colhead{$\frac{M_{\rm ic}}{M_\sun}$} & \colhead{$\frac{\psi_{\rm c}}{kT}$} & \colhead{$\log_{10} (\frac{\rho_{\rm c}}{{\rm g\:cm^{-3}}})$}
		& \colhead{$\log_{10} (\frac{T_{\rm c}}{{\rm K}})$} & \colhead{$X_{\rm c}$} & \colhead{$Y_{\rm c}$} & \colhead{$X_{\rm s}$} }
	\startdata
	\cutinhead{\normalsize $3.2000\ M_\sun$}
1	&	6.2080	&	0.0000	&	0.8850	&	0.0000	&	-3.465	&	1.546	&	7.383	&	0.696	&	0.283	&	0.699	\\*
2	&	7.9080	&	0.0000	&	0.8474	&	0.0000	&	-3.580	&	1.528	&	7.386	&	0.597	&	0.383	&	0.699	\\
3	&	8.1498	&	0.0000	&	0.8102	&	0.0000	&	-3.677	&	1.525	&	7.394	&	0.499	&	0.482	&	0.699	\\
4	&	8.2762	&	0.0000	&	0.7729	&	0.0000	&	-3.762	&	1.531	&	7.403	&	0.401	&	0.579	&	0.699	\\
5	&	8.3595	&	0.0000	&	0.7337	&	0.0000	&	-3.841	&	1.548	&	7.415	&	0.300	&	0.681	&	0.699	\\
6	&	8.4142	&	0.0000	&	0.6958	&	0.0000	&	-3.900	&	1.577	&	7.429	&	0.202	&	0.779	&	0.699	\\
7	&	8.4533	&	0.0000	&	0.6578	&	0.0000	&	-3.930	&	1.630	&	7.449	&	0.104	&	0.877	&	0.699	\\
8	&	8.4765	&	0.0000	&	0.6271	&	0.0000	&	-3.885	&	1.736	&	7.485	&	0.026	&	0.954	&	0.699	\\
9	&	8.4827	&	0.0000	&	0.6162	&	0.0000	&	-2.736	&	2.336	&	7.549	&	0.000	&	0.980	&	0.699	\\
10	&	8.4830	&	0.0000	&	0.6157	&	0.0000	&	-1.794	&	2.684	&	7.518	&	0.000	&	0.980	&	0.699	\\
11	&	8.4838	&	0.0000	&	0.6154	&	0.0000	&	-1.100	&	2.981	&	7.530	&	0.000	&	0.980	&	0.699	\\
12	&	8.4846	&	0.0000	&	0.6152	&	0.0000	&	-0.663	&	3.214	&	7.573	&	0.000	&	0.980	&	0.699	\\
13	&	8.4851	&	0.0000	&	0.6151	&	0.0000	&	-0.403	&	3.377	&	7.616	&	0.000	&	0.980	&	0.698	\\
14	&	8.4856	&	0.0047	&	0.6149	&	0.0000	&	-0.120	&	3.572	&	7.678	&	0.000	&	0.980	&	0.698	\\
15	&	8.4859	&	0.1401	&	0.6147	&	0.0000	&	0.043	&	3.690	&	7.719	&	0.000	&	0.980	&	0.698	\\
16	&	8.4860	&	0.3395	&	0.6146	&	0.0000	&	0.114	&	3.742	&	7.737	&	0.000	&	0.980	&	0.698	\\
17	&	8.4864	&	1.2953	&	0.6141	&	0.0000	&	0.347	&	3.902	&	7.792	&	0.000	&	0.980	&	0.698	\\
18	&	8.4868	&	1.9842	&	0.6133	&	0.0000	&	0.636	&	4.084	&	7.851	&	0.000	&	0.980	&	0.697	\\
19	&	8.4873	&	2.3994	&	0.6092	&	0.0000	&	0.953	&	4.272	&	7.912	&	0.000	&	0.980	&	0.688	\\
20	&	8.4878	&	2.5843	&	0.5994	&	0.0000	&	1.276	&	4.455	&	7.973	&	0.000	&	0.980	&	0.673	\\
21	&	8.4883	&	2.6446	&	0.5488	&	0.0912	&	1.035	&	4.535	&	8.068	&	0.000	&	0.979	&	0.665	\\
22	&	8.5045	&	0.8827	&	0.5559	&	0.2191	&	-0.126	&	4.196	&	8.088	&	0.000	&	0.833	&	0.665	\\
23	&	8.5652	&	0.5946	&	0.6420	&	0.3891	&	-0.771	&	4.056	&	8.152	&	0.000	&	0.255	&	0.665	\\
24	&	8.5848	&	0.9931	&	0.6675	&	0.5042	&	-1.099	&	3.948	&	8.164	&	0.000	&	0.134	&	0.665	\\
25	&	8.5850	&	0.7203	&	0.6677	&	0.5146	&	-0.958	&	4.019	&	8.174	&	0.000	&	0.128	&	0.665	\\
26	&	8.5893	&	1.4878	&	0.6730	&	0.5089	&	-0.668	&	4.269	&	8.264	&	0.000	&	0.011	&	0.665	\\
27	&	8.5900	&	2.0177	&	0.6743	&	0.5076	&	-0.282	&	4.511	&	8.327	&	0.000	&	0.000	&	0.665	\\
28	&	8.5901	&	2.2555	&	0.6749	&	0.5068	&	0.524	&	4.799	&	8.336	&	0.000	&	0.000	&	0.665	\\
29	&	8.5903	&	2.1974	&	0.6753	&	0.5054	&	1.236	&	4.966	&	8.308	&	0.000	&	0.000	&	0.665	\\
30	&	8.5910	&	2.3732	&	0.6759	&	0.5051	&	2.563	&	5.300	&	8.320	&	0.000	&	0.000	&	0.665	\\
31	&	8.5913	&	2.4422	&	0.6760	&	0.5080	&	3.157	&	5.455	&	8.345	&	0.000	&	0.000	&	0.665	\\
32	&	8.5916	&	2.4787	&	0.6760	&	0.5278	&	4.021	&	5.648	&	8.375	&	0.000	&	0.000	&	0.665	\\
33	&	8.5918	&	2.4978	&	0.6761	&	0.5467	&	4.952	&	5.773	&	8.370	&	0.000	&	0.000	&	0.665	\\
34	&	8.5920	&	2.5098	&	0.6761	&	0.5718	&	6.542	&	5.922	&	8.347	&	0.000	&	0.000	&	0.665	\\
35	&	8.5921	&	2.5156	&	0.6761	&	0.5993	&	8.761	&	6.065	&	8.311	&	0.000	&	0.000	&	0.665	\\
36	&	8.5922	&	2.5179	&	0.6761	&	0.6275	&	11.893	&	6.204	&	8.262	&	0.000	&	0.000	&	0.664	\\
37	&	8.5923	&	2.5184	&	0.6767	&	0.6538	&	16.618	&	6.346	&	8.200	&	0.000	&	0.000	&	0.664	\\
38	&	8.5925	&	2.5109	&	0.6874	&	0.6769	&	26.112	&	6.519	&	8.100	&	0.000	&	0.000	&	0.664	\\
39	&	8.5927	&	2.4887	&	0.7105	&	0.7044	&	38.329	&	6.663	&	8.012	&	0.000	&	0.000	&	0.664	\\
40	&	8.5930	&	2.4352	&	0.7645	&	0.7611	&	54.312	&	6.844	&	7.956	&	0.000	&	0.000	&	0.664	\\
41	&	8.5933	&	2.3695	&	0.8304	&	0.8284	&	67.707	&	7.021	&	7.950	&	0.000	&	0.000	&	0.664	\\
42	&	8.5935	&	2.3217	&	0.8783	&	0.8769	&	76.247	&	7.145	&	7.960	&	0.000	&	0.000	&	0.664	\\
43	&	8.5936	&	2.2723	&	0.9277	&	0.9267	&	84.729	&	7.276	&	7.977	&	0.000	&	0.000	&	0.664	\\
44	&	8.5938	&	2.2072	&	0.9928	&	0.9922	&	96.061	&	7.454	&	8.006	&	0.000	&	0.000	&	0.664	\\
45	&	8.5940	&	2.1427	&	1.0573	&	1.0569	&	108.211	&	7.645	&	8.040	&	0.000	&	0.000	&	0.664	\\
46	&	8.5941	&	2.0504	&	1.1496	&	1.1494	&	128.359	&	7.955	&	8.100	&	0.000	&	0.000	&	0.664	\\
47	&	8.5943	&	1.9444	&	1.2556	&	1.2555	&	158.489	&	8.419	&	8.198	&	0.000	&	0.000	&	0.664	\\
48	&	8.5945	&	1.8282	&	1.3718	&	1.3718	&	203.187	&	9.363	&	8.449	&	0.000	&	0.000	&	0.664	\\
	\enddata
	
	\tablecomments{Table \ref{intmod} is published in its entirety in the electronic edition
		of the {\it Astrophysical Journal Supplement}.  A portion is shown here for guidance regarding its form and content.}
	
\end{deluxetable}

\newcounter{tbl1}

\begin{deluxetable}{rrrrrrrrrr}    
	\tabletypesize{\footnotesize}
	\tablewidth{0pt}
	\tablecolumns{10}
	\tablecaption{Global properties of initial models\label{glbmod}}
	
	\tablehead{
		\colhead{$k$} & \colhead{$\log_{10} (\frac{R}{R_\sun})$} & \colhead{$\log_{10} (\frac{T_{\rm e}}{{\rm K}})$} & \colhead{$\log_{10} (\frac{L}{L_\sun})$} &
		\colhead{$\log_{10} (\frac{L_{\rm H}}{L_\sun})$} & \colhead{$\log_{10} {\frac{L_{\rm He}}{L_\sun}}$} & \colhead{$\log{10} (\frac{L_Z}{L_\sun})$} &
		\colhead{$\log_{10} (\frac{|L_\nu|}{L_\sun})$} & \colhead{$\log_{10} (\frac{|L_{\rm th}|}{L_\sun})$} & \colhead{$\frac{I}{MR^2}$} 
	}
	\startdata
	\cutinhead{\normalsize $3.2000\ M_\sun$}
1	&	0.3185	&	4.1067	&	2.0169	&	2.046	&	-24.817	&	\nodata\phn	&	0.850*	&	-0.742*	&	0.0532	\\*
2	&	0.3701	&	4.0943	&	2.0706	&	2.100	&	-24.468	&	\nodata\phn	&	0.907*	&	-0.856*	&	0.0484	\\
3	&	0.4211	&	4.0816	&	2.1220	&	2.151	&	-24.069	&	\nodata\phn	&	0.960*	&	-0.794*	&	0.0441	\\
4	&	0.4749	&	4.0662	&	2.1680	&	2.197	&	-23.649	&	\nodata\phn	&	1.004*	&	-1.607	&	0.0401	\\
5	&	0.5367	&	4.0458	&	2.2099	&	2.239	&	-23.168	&	\nodata\phn	&	1.045*	&	-1.491	&	0.0363	\\
6	&	0.6028	&	4.0209	&	2.2426	&	2.271	&	-22.623	&	\nodata\phn	&	1.076*	&	-1.353	&	0.0330	\\
7	&	0.6744	&	3.9914	&	2.2677	&	2.296	&	-21.865	&	\nodata\phn	&	1.100*	&	-0.926	&	0.0301	\\
8	&	0.7196	&	3.9755	&	2.2946	&	2.322	&	-20.517	&	\nodata\phn	&	1.126*	&	-0.339	&	0.0281	\\
9	&	0.6531	&	4.0319	&	2.3872	&	2.457	&	-17.617	&	\nodata\phn	&	1.261*	&	1.381*	&	0.0272	\\
10	&	0.7516	&	3.9852	&	2.3971	&	2.459	&	-18.162	&	\nodata\phn	&	1.264*	&	1.303*	&	0.0245	\\
11	&	0.8538	&	3.9355	&	2.4027	&	2.452	&	-17.557	&	\nodata\phn	&	1.256*	&	1.087*	&	0.0222	\\
12	&	0.9534	&	3.8790	&	2.3763	&	2.436	&	-16.111	&	\nodata\phn	&	1.240*	&	1.241*	&	0.0209	\\
13	&	1.0493	&	3.8199	&	2.3313	&	2.410	&	-14.687	&	\nodata\phn	&	1.214*	&	1.416*	&	0.0203	\\
14	&	1.1532	&	3.7380	&	2.2118	&	2.343	&	-12.734	&	\nodata\phn	&	1.148*	&	1.639*	&	0.0281	\\
15	&	1.1434	&	3.7163	&	2.1053	&	2.281	&	-11.489	&	\nodata\phn	&	1.086*	&	1.712*	&	0.0592	\\
16	&	1.1516	&	3.7078	&	2.0877	&	2.256	&	-10.951	&	\nodata\phn	&	1.061*	&	1.668*	&	0.0786	\\
17	&	1.2551	&	3.6859	&	2.2070	&	2.262	&	-9.346	&	\nodata\phn	&	1.067*	&	1.009*	&	0.1163	\\
18	&	1.3566	&	3.6719	&	2.3542	&	2.376	&	-7.554	&	\nodata\phn	&	1.181*	&	0.538	&	0.1266	\\
19	&	1.4598	&	3.6592	&	2.5096	&	2.524	&	-5.263	&	\nodata\phn	&	1.329*	&	1.017	&	0.1313	\\
20	&	1.5650	&	3.6465	&	2.6692	&	2.682	&	-2.223	&	\nodata\phn	&	1.486*	&	1.232	&	0.1348	\\
21	&	1.6659	&	3.6338	&	2.8203	&	2.831	&	1.541	&	-35.172	&	1.636*	&	0.906*	&	0.1353	\\
22	&	1.2213	&	3.6949	&	2.1755	&	2.096	&	1.517	&	-30.489	&	0.907*	&	-0.844	&	0.1067	\\
23	&	1.2919	&	3.6923	&	2.3061	&	2.064	&	1.975	&	-26.020	&	0.885*	&	-0.693*	&	0.0836	\\
24	&	1.3626	&	3.6804	&	2.4001	&	1.978	&	2.279	&	-25.925	&	0.809*	&	1.437*	&	0.0956	\\
25	&	1.3140	&	3.6883	&	2.3344	&	1.974	&	2.065	&	-25.309	&	0.808*	&	1.086	&	0.0882	\\
26	&	1.3971	&	3.6729	&	2.4389	&	2.100	&	2.188	&	-20.796	&	0.990*	&	0.644	&	0.1107	\\
27	&	1.4840	&	3.6604	&	2.5630	&	2.354	&	1.841	&	-17.701	&	1.276*	&	1.950	&	0.1186	\\
28	&	1.5684	&	3.6492	&	2.6868	&	2.453	&	2.221	&	-16.811	&	1.394*	&	1.786	&	0.1202	\\
29	&	1.5424	&	3.6526	&	2.6486	&	2.143	&	2.477	&	-17.801	&	1.167*	&	1.320	&	0.1199	\\
30	&	1.6412	&	3.6394	&	2.7934	&	1.143	&	2.767	&	-16.494	&	1.057*	&	1.537	&	0.1211	\\
31	&	1.7362	&	3.6268	&	2.9328	&	0.778	&	2.896	&	-14.985	&	1.255*	&	1.909	&	0.1208	\\
32	&	1.8392	&	3.6129	&	3.0831	&	1.106	&	3.061	&	-12.964	&	1.581*	&	1.938	&	0.1197	\\
33	&	1.9344	&	3.5999	&	3.2215	&	0.648	&	3.203	&	-12.489	&	1.706*	&	2.065	&	0.1185	\\
34	&	2.0382	&	3.5855	&	3.3719	&	0.509	&	3.350	&	-12.058	&	1.874*	&	2.271	&	0.1172	\\
35	&	2.1338	&	3.5723	&	3.5099	&	1.309	&	3.486	&	-11.376	&	2.063*	&	2.433	&	0.1164	\\
36	&	2.2170	&	3.5607	&	3.6301	&	2.651	&	3.697	&	-10.605	&	2.296*	&	2.983*	&	0.1164	\\
37	&	2.3140	&	3.5472	&	3.7700	&	3.421	&	3.490	&	-9.932	&	2.599*	&	2.748	&	0.1176	\\
38	&	2.4199	&	3.5325	&	3.9229	&	3.840	&	3.224	&	-10.439	&	2.816*	&	2.634	&	0.1210	\\
39	&	2.5159	&	3.5189	&	4.0604	&	4.001	&	3.534	&	-11.896	&	2.887*	&	3.067*	&	0.1266	\\
40	&	2.6175	&	3.5037	&	4.2030	&	4.153	&	3.369	&	-12.640	&	2.999*	&	2.606	&	0.1354	\\
41	&	2.7034	&	3.4899	&	4.3195	&	4.270	&	3.474	&	-11.896	&	3.109*	&	2.756	&	0.1457	\\
42	&	2.7588	&	3.4806	&	4.3934	&	4.343	&	3.546	&	-11.146	&	3.182*	&	2.852	&	0.1543	\\
43	&	2.8134	&	3.4716	&	4.4663	&	4.415	&	3.636	&	-10.315	&	3.255*	&	2.859	&	0.1643	\\
44	&	2.8694	&	3.4629	&	4.5436	&	4.491	&	3.725	&	-9.318	&	3.333*	&	2.913	&	0.1752	\\
45	&	2.9067	&	3.4580	&	4.5985	&	4.544	&	3.737	&	-8.531	&	3.389*	&	3.221	&	0.1820	\\
46	&	2.9422	&	3.4546	&	4.6560	&	4.597	&	3.816	&	-7.575	&	3.446*	&	3.297	&	0.1869	\\
47	&	2.9704	&	3.4535	&	4.7080	&	4.640	&	3.828	&	-6.434	&	3.496*	&	3.578	&	0.1887	\\
48	&	2.9993	&	3.4538	&	4.7669	&	4.665	&	3.838	&	1.351	&	3.566*	&	3.956	&	0.1899	\\
	\enddata
	
	\tablecomments{Table \ref{glbmod} is published in its entirety in the electronic edition of the {\it Astrophysical Journal Supplement}. A portion is shown here for guidance regarding its form and content.}
	\tablecomments{The asterisk, *, is appended to signify that the neurino or gravothermal luminosity is negative.}
	
\end{deluxetable}

\newcounter{tbl2}

\begin{list}{(\arabic{tbl1})}{\usecounter{tbl1}}
	\item $k$ --- mass loss sequence number,
	\item $t$ --- age,
	\item $M_{\rm ce}$ --- mass of the convective envelope,
	\item $M_{\rm c}$ --- core mass,
	\item $M_{\rm ic}$ --- inner core mass,
	\item $\psi_{\rm c}$ --- central electron chemical potential ($\mu_e$),
	\item $\rho_{\rm c}$ --- central density,
	\item $T_{\rm c}$ --- central temperature,
	\item $X_{\rm c}$ --- central hydrogen abundance (fraction by mass),
	\item $Y_{\rm c}$ --- central helium abundance (fraction by mass) and
	\item $X_{\rm s}$ --- surface hydrogen abundance (fraction by mass).
\end{list}

Age $t$ is measured from the ZAMS model (excluding pre-main-sequence evolution). The mass of the convective envelope $M_{\rm ce}$ refers to the mass depth of the base of the outermost convection zone. The core mass $M_{\rm c}$ refers to the mass coordinate at which the helium abundance is halfway between the surface helium abundance and the maximum helium abundance in the stellar interior. The inner core mass $M_{\rm ic}$ identifies the mass coordinate at which the helium abundance is halfway between the maximum helium abundance in the stellar interior and the minimum helium abundance interior to that maximum. In the absence of measurable helium depletion in the hydrogen-exhaused core, $M_{\rm ic}$ is set to a default value of zero. $M_{\rm c}$ and $M_{\rm ic}$ characterize the \emph{range} in mass over which hydrogen and helium are being depleted during their respective core burning phases, and \emph{not} the amount of mass that has been consumed. Upon core fuel exhaustion, $M_{\rm c}$ and $M_{\rm ic}$ mark the midpoints in hydrogen and helium depletion profiles, respectively. The dimensionless central electron chemical potential $\psi_{\rm c}$ measures the degree of electron degeneracy (once $\psi_{\rm c} > 0$).

Like Table~\ref{intmod}, Table~\ref{glbmod} is arranged in segments, by stellar mass, $M_{\rm i}$. Successive columns list

\begin{list}{(\arabic{tbl2})}{\usecounter{tbl2}}
	\item $k$ --- mass loss sequence number,
	\item $R$ --- radius,
	\item $T_e$ --- effective temperature,
	\item $L$ --- stellar luminosity,
	\item $L_{\rm H}$ --- hydrogen-burning luminosity,
	\item $L_{\rm He}$ --- helium-burning luminosity,
	\item $L_Z$ --- heavy-element (carbon-, oxygen-, etc.) burning luminosity,
	\item $|L_{\nu}|$ --- log neutrino luminosity (with asterisk, *, appended to signify that this is a \emph{negative} contribution to the net stellar luminosity),
	\item $|L_{\rm th}|$ --- gravothermal luminosity (with asterisk, *,
	appended where the gravothermal luminosity is negative) and
	\item $I/(MR^2)$ --- dimensionless moment of inertia.
\end{list}

Table~\ref{theqml} summaries the quantitative results of our investigation for those initial stellar model sequences. For each set of sequences, it identifies critical points marking the onset of thermal timescale mass transfer or unstable mass transfer for overfilling outer Lagrangian surface, and the (critical) initial conditions (radius-mass exponent and mass ratio) corresponding to those critical points.

Table~\ref{theqml} is arranged in segments, by stellar mass of the donor, $M_{\rm i}$. Successive columns list

\newcounter{tbl3}
\begin{list}{(\arabic{tbl3})}{\usecounter{tbl3}}
	\item $k$ --- mass-loss sequence number,
    \item $R_{\rm i}$ --- initial radius,
	\item $M_{\rm KH}$ --- mass threshold at which $\dot{M} = - M_{\rm i}/\tau^{\rm i}_{\rm KH}$,
	\item $R_{\rm KH}$ --- Roche-lobe radius at which $\dot{M} = - M_{\rm i}/\tau^{\rm i}_{\rm KH}$,
	\item $R_{\rm KH}^*$ --- stellar radius when $\dot{M} = - M_{\rm i}/\tau^{\rm i}_{\rm KH}$,
	\item $\zeta_{\rm th}$ --- critical radius-mass exponent for thermal timescale mass transfer,
	\item $q_{\rm th}$ --- critical mass ratio for thermal timescale (conservative) mass transfer,
	\item $M_{\rm L_2}$ --- mass threshold at outer Lagrangian point ${\rm L_2}$,
	\item $R_{\rm L_2}$ --- Roche-lobe radius at outer Lagrangian point ${\rm L_2}$,
	\item $\zeta_{\rm L_2}$ --- critical radius-mass exponent for unstable mass transfer overfilling outer Lagrangian point 
	and
	\item $q_{\rm L_2}$ --- critical mass ratio for unstable (conservative) mass transfer overfilling outer Lagrangian point.

\end{list}

\begin{deluxetable}{crrrrrrrrrr}
	\tabletypesize{\footnotesize}
    \tablewidth{0pt}
	\tablecolumns{11}
	\tablecaption{Thresholds for conservative thermal timescale mass transfer\label{theqml}}
	
	\tablehead{
	    \colhead{$k$} & \colhead{$\log_{10} (\frac{R_{\rm i}}{R_\sun})$} & \colhead{$\frac{M_{\rm KH}}{M_\sun}$} & \colhead{$\log_{10} (\frac{R_{\rm KH}}{R_\sun})$} & \colhead{$\log_{10} (\frac{R^*_{\rm KH}}{R_\sun})$} & \colhead{$\zeta_{\rm th}$} & \colhead{$q_{\rm th}$} & \colhead{$\frac{M_{\rm L_2}}{M_\sun}$} & \colhead{$\log_{10} (\frac{R_{\rm L_2}}{R_\sun})$} & \colhead{$\zeta_{\rm L_2}$} & \colhead{$q_{\rm L_2}$} }
	\startdata
	\cutinhead{\normalsize $3.2000\ M_\sun$}
1	&	0.3185	&	2.9405	&	0.2908	&	0.2986	&	0.985	&	1.245	&	2.5943	&	0.2696	&	2.741	&	2.063	\\
2	&	0.3701	&	2.9213	&	0.3431	&	0.3518	&	0.921	&	1.215	&	2.5794	&	0.3272	&	2.641	&	2.016	\\
3	&	0.4211	&	2.8968	&	0.3958	&	0.4056	&	0.831	&	1.174	&	2.5542	&	0.3865	&	2.507	&	1.953	\\
4	&	0.4749	&	2.8619	&	0.4529	&	0.4638	&	0.709	&	1.117	&	2.2667	&	0.4438	&	2.269	&	1.843	\\
5	&	0.5368	&	2.8047	&	0.5215	&	0.5339	&	0.529	&	1.033	&	2.1679	&	0.5363	&	1.894	&	1.668	\\
6	&	0.6028	&	2.6978	&	0.6031	&	0.6169	&	0.275	&	0.916	&	1.9476	&	0.6688	&	1.340	&	1.410	\\
7	&	0.6745	&	2.1205	&	0.7842	&	0.7957	&	-0.227	&	0.683	&	2.0919	&	0.7994	&	0.639	&	1.085	\\
8	&	0.7197	&	2.2672	&	0.8703	&	0.8797	&	-0.834	&	0.402	&	2.2480	&	0.8825	&	0.083	&	0.827	\\
9	&	0.6533	&	1.8187	&	0.7741	&	0.7851	&	0.160	&	0.862	&	1.8280	&	0.7837	&	0.902	&	1.207	\\
10	&	0.7518	&	2.4539	&	0.9469	&	0.9596	&	-1.675	&	-0.002	&	2.3372	&	1.0020	&	-1.049	&	0.302	\\
11	&	0.8540	&	3.0963	&	0.8798	&	0.8960	&	-1.689	&	-0.013	&	2.9026	&	1.0017	&	-1.727	&	-0.042	\\
12	&	0.9536	&	3.1608	&	0.9659	&	0.9824	&	-1.706	&	-0.026	&	3.0502	&	1.0931	&	-1.734	&	-0.046	\\
13	&	1.0493	&	3.1840	&	1.0559	&	1.0710	&	-1.757	&	-0.063	&	2.7246	&	1.1537	&	0.456	&	1.000	\\
14	&	1.1528	&	3.1993	&	1.1394	&	1.1500	&	152.024	&	74.458	&	-9.9900	&	-9.9900	&	-9.990	&	-9.990	\\
15	&	1.1434	&	3.1677	&	1.1295	&	1.1357	&	3.252	&	2.301	&	3.1593	&	1.1307	&	9.421	&	5.204	\\
16	&	1.1516	&	2.9528	&	1.1594	&	1.1636	&	-0.118	&	0.734	&	2.1462	&	1.2063	&	1.303	&	1.392	\\
17	&	1.2552	&	2.9030	&	1.2568	&	1.2622	&	0.110	&	0.839	&	2.1577	&	1.2849	&	1.564	&	1.514	\\
18	&	1.3566	&	2.8962	&	1.3574	&	1.3644	&	0.137	&	0.852	&	2.1635	&	1.3873	&	1.551	&	1.508	\\
19	&	1.4598	&	2.8565	&	1.4608	&	1.4700	&	0.159	&	0.862	&	2.1385	&	1.4959	&	1.503	&	1.486	\\
20	&	1.5650	&	2.7787	&	1.5671	&	1.5795	&	0.188	&	0.875	&	2.1242	&	1.6063	&	1.454	&	1.463	\\
21	&	1.6659	&	2.7388	&	1.6660	&	1.6825	&	0.251	&	0.904	&	2.1346	&	1.7087	&	1.434	&	1.454	\\
22	&	1.2213	&	2.9453	&	1.2100	&	1.2150	&	0.482	&	1.012	&	2.2026	&	1.2071	&	2.051	&	1.741	\\
23	&	1.2919	&	2.8852	&	1.2756	&	1.2820	&	0.585	&	1.059	&	2.1969	&	1.2726	&	2.111	&	1.769	\\
24	&	1.3626	&	2.8967	&	1.3514	&	1.3591	&	0.454	&	0.999	&	2.2371	&	1.3521	&	2.011	&	1.722	\\
25	&	1.3140	&	2.8820	&	1.3020	&	1.3088	&	0.472	&	1.007	&	2.1693	&	1.3108	&	1.925	&	1.682	\\
26	&	1.3971	&	2.8845	&	1.3919	&	1.4000	&	0.297	&	0.926	&	2.1841	&	1.4104	&	1.737	&	1.594	\\
27	&	1.4839	&	2.8639	&	1.4815	&	1.4917	&	0.237	&	0.898	&	2.1618	&	1.5110	&	1.593	&	1.527	\\
28	&	1.5682	&	2.7762	&	1.5651	&	1.5791	&	0.295	&	0.925	&	2.1758	&	1.5971	&	1.569	&	1.516	\\
29	&	1.5422	&	2.7851	&	1.5401	&	1.5530	&	0.269	&	0.913	&	2.1448	&	1.5730	&	1.560	&	1.512	\\
30	&	1.6411	&	2.7286	&	1.6422	&	1.6591	&	0.243	&	0.901	&	2.1624	&	1.6835	&	1.425	&	1.450	\\
31	&	1.7359	&	2.7222	&	1.7332	&	1.7553	&	0.318	&	0.935	&	2.1613	&	1.7796	&	1.415	&	1.445	\\
32	&	1.8392	&	2.6167	&	1.8324	&	1.8627	&	0.435	&	0.990	&	2.1150	&	1.8865	&	1.397	&	1.436	\\
33	&	1.9346	&	2.5813	&	1.9197	&	1.9598	&	0.568	&	1.052	&	2.1093	&	1.9828	&	1.390	&	1.433	\\
34	&	2.0384	&	2.4959	&	2.0132	&	2.0679	&	0.744	&	1.133	&	2.1120	&	2.0870	&	1.384	&	1.430	\\
35	&	2.1341	&	2.4566	&	2.0936	&	2.1653	&	0.951	&	1.229	&	2.1093	&	2.1829	&	1.381	&	1.429	\\
36	&	2.2172	&	2.3965	&	2.1584	&	2.2512	&	1.195	&	1.343	&	2.1111	&	2.2658	&	1.384	&	1.430	\\
37	&	2.3138	&	2.3158	&	2.2176	&	2.3523	&	1.657	&	1.557	&	2.1123	&	2.3633	&	1.378	&	1.427	\\
38	&	2.4199	&	2.2876	&	2.2357	&	2.4626	&	2.727	&	2.056	&	2.0775	&	2.4759	&	1.331	&	1.406	\\
39	&	2.5175	&	2.3189	&	2.1618	&	2.5549	&	5.188	&	3.208	&	2.0823	&	2.6078	&	0.970	&	1.238	\\
40	&	2.6173	&	2.3070	&	0.2085	&	2.6331	&	124.019	&	60.742	&	2.0703	&	2.6882	&	1.190	&	1.340	\\
41	&	2.7032	&	2.1779	&	-0.5246	&	2.7214	&	293.101	&	143.828	&	1.9844	&	2.7358	&	1.633	&	1.546	\\
42	&	2.7587	&	2.4556	&	-1.0347	&	2.7876	&	704.588	&	347.274	&	2.4127	&	2.8303	&	1.008	&	1.256	\\
43	&	2.8134	&	2.4079	&	-1.6015	&	2.8275	&	1361.881	&	673.436	&	2.3480	&	2.8321	&	1.675	&	1.566	\\
44	&	2.8693	&	\nodata\phn	&	\nodata\phn	&	\nodata\phn	&	\nodata\phn	&	\nodata\phn	&	2.3746	&	2.8675	&	1.951	&	1.694	\\
45	&	2.9066	&	\nodata\phn	&	\nodata\phn	&	\nodata\phn	&	\nodata\phn	&	\nodata\phn	&	2.7515	&	2.8957	&	2.412	&	1.909	\\
46	&	2.9422	&	\nodata\phn	&	\nodata\phn	&	\nodata\phn	&	\nodata\phn	&	\nodata\phn	&	3.0884	&	2.9382	&	4.595	&	2.930	\\
47	&	2.9704	&	\nodata\phn	&	\nodata\phn	&	\nodata\phn	&	\nodata\phn	&	\nodata\phn	&	\nodata\phn	&	\nodata\phn	&	\nodata\phn	&	\nodata\phn	\\
48	&	2.9993	&	\nodata\phn	&	\nodata\phn	&	\nodata\phn	&	\nodata\phn	&	\nodata\phn	&	\nodata\phn	&	\nodata\phn	&	\nodata\phn	&	\nodata\phn	\\*
	\enddata
	
	\tablecomments{Table \ref{theqml} is published in its entirety in the electronic edition of the {\it Astrophysical Journal Supplement}.  A portion is shown here for guidance regarding its form and content.}
	
\end{deluxetable}

We list a portion of our results, a sequence of $3.2~M_\sun$ donor stars from ZAMS to TPAGB, for thermal timescale mass transfer and unstable mass transfer by overfilling the outer Lagrangian surface in Table \ref{theqml}. The methods to constrain the possible mass transfer channels are given in Section~\ref{sec_mt3_2}.

\begin{figure}[ht!]
	\centering
	\includegraphics[scale=0.4]{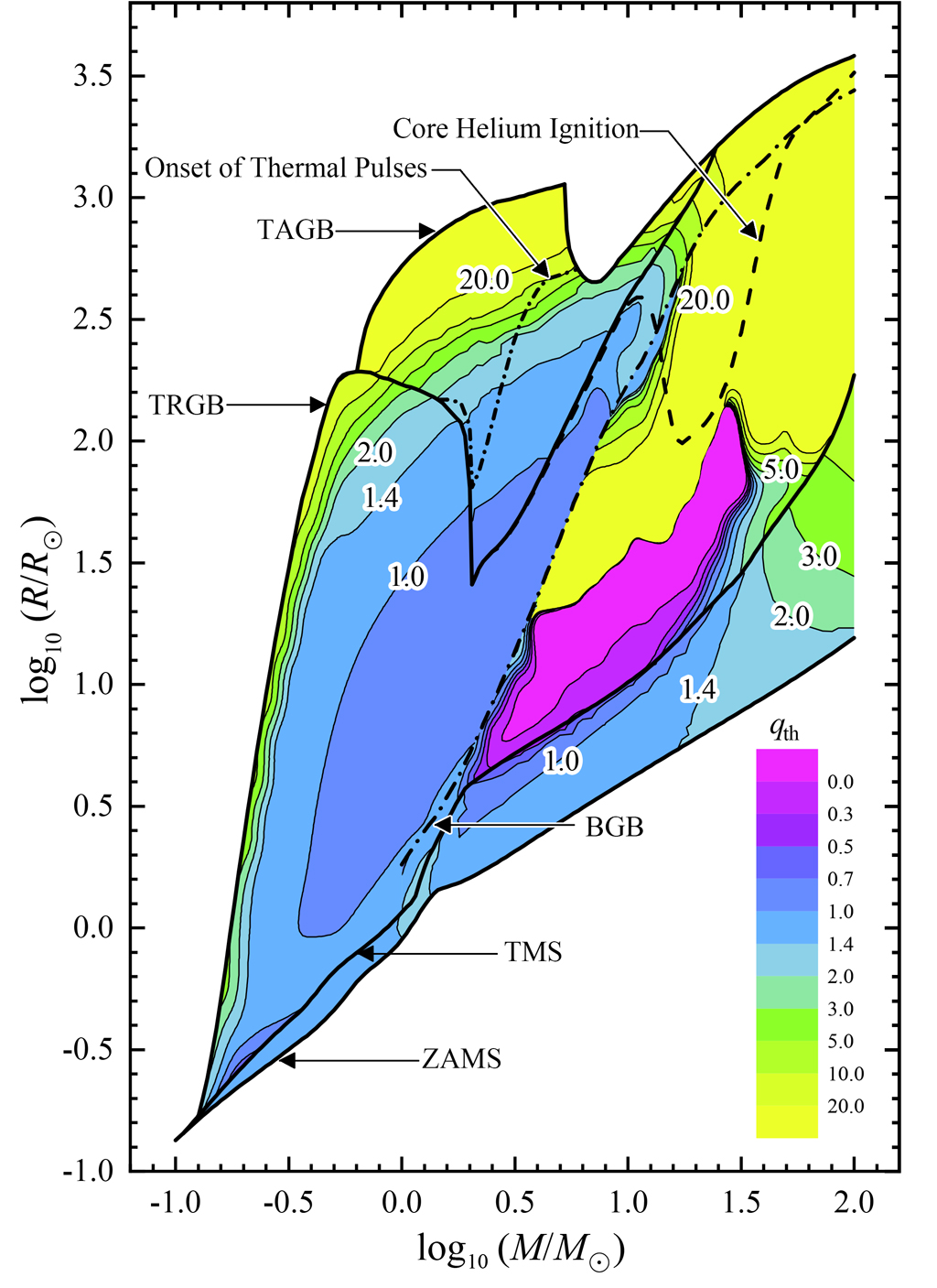}
	\caption{Contour map of the critical mass ratio $q_{\rm th}$ for thermal timescale mass transfer on the mass-radius diagram (MRD). The first thick solid black line, counting from the bottom, shows the masses and radii of ZAMS stars. The second thick solid black line marks the masses and radii of the TMS stars. The third thick solid black line shows the tip of first giant branch stars (or RGB for low- and intermediate-mass stars). The fourth thick solid black line is the tip of AGB stars or the maximum radius of massive stars. The long dash-dotted line indicates stars at the base of the giant branch. The dashed line marks the starting position of the core helium-burning phase. The short dash-dotted line shows the position of donor stars, which begin to pulsate thermally on AGB. \label{fig07}}
\end{figure}

\begin{figure}[ht!]
	\centering
	\includegraphics[scale=0.4]{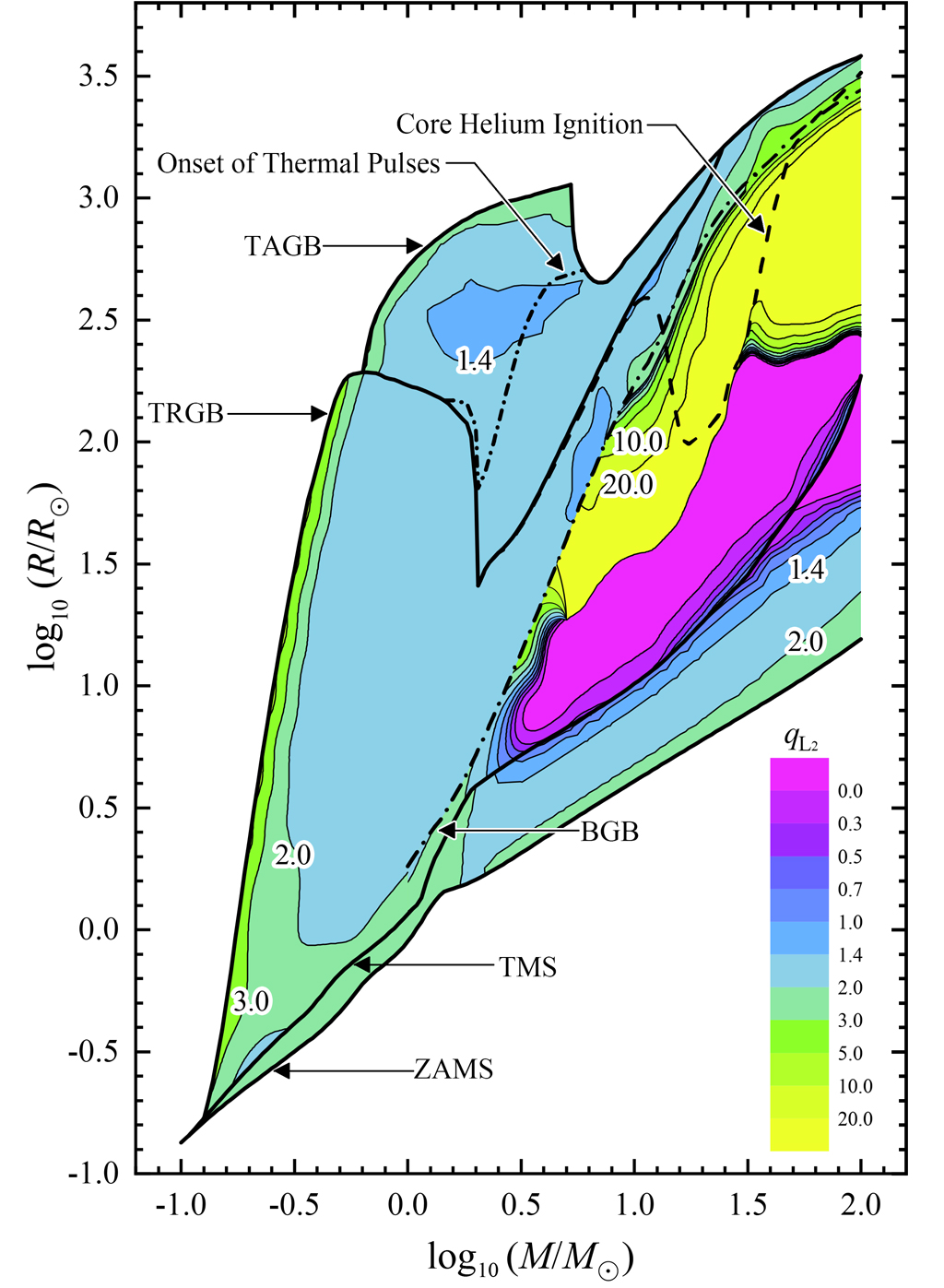}
	\caption{Contour map of the critical mass ratio $q_{\rm L_2}$ for donor stars overfilling their outer Lagrangian surface $R_{\rm L_2}$ (another kind of unstable mass transfer to form a common envelope, CE) on the MRD. Different thick black lines have the same meanings as explained in Figure~\ref{fig07}. \label{fig08}}
\end{figure}

Figures~\ref{fig07} and~\ref{fig08} show contour maps of the threshold mass ratios for conservative thermal timescale mass transfer and unstable mass transfer through overfilling the outer Lagrangian surface, respectively, on a mass--radius diagram (MRD). All initial donor model sequences cover from $0.1$ to $100~M_{\odot}$, except the radius shadow region (in which the radius shrinks compared with its previous evolutionary stage. That is we omit the core helium burning models after the tip of RGB models for the low- and intermediate-mass stars).

First we discuss the thresholds for donor stars on the MS in Figures~\ref{fig07} and~\ref{fig08} between the first (ZAMS stars) and second (TMS stars) thick solid black lines counting from the bottom. Just like the $3.2~M_\odot$ donor star, stars on the MS with a mass larger than $1.6~M_\odot$ become less stable ($q_{\rm th}$ decreases) for thermal timescale mass transfer as the stars evolve due to the slight expansion of the envelope and the shrinkage of the nuclear burning core. With increasing of donor mass, $q_{\rm th}$ becomes larger and larger as the thermal timescale decreases from around $10^6$ years to about $10^4$ years. The critical mass ratio for thermal timescale mass transfer of low-mass MS donor stars with masses less than $1.0~M_\odot$ changes slightly and not obviously between 1.4 and 1.0 in Figure~\ref{fig07}. The critical mass ratio $q_{\rm L_2}$ for unstable overflow through the outer Lagrangian radius has the same pattern as $q_{\rm th}$, but to a greater extent (Figure~\ref{fig08}).

Secondly we study the thresholds of donor stars on the HG in Figures~\ref{fig07} and~\ref{fig08} between the second thick solid black line (TMS) counting from the bottom and the long-dash-dotted line (base of giant branch stars, or BGB stars). Donor stars on the early stage of the HG, similar to the $3.2~M_\odot$ donor star, expand on a thermal timescale and shrink suddenly at the late stages of the HG. So both $q_{\rm th}$ and $q_{\rm L_2}$ decrease to a minimum and then suddenly jump up considerably.

Finally, the thresholds of donor stars on the RGB and AGB are shown in Figures~\ref{fig07} and~\ref{fig08} between the long-dash-dotted line (BGB stars) and the fourth thick solid black line (tip of AGB stars or massive stars with maximum radius) counting from the bottom. The critical mass ratio $q_{\rm th}$ for thermal timescale mass transfer increases with increasing/decreasing mass (diverging at around $1.0~M_\odot$) and increases at more evolved stages. The thermal timescale of these donor stars drops as they evolve. The critical mass ratios $q_{\rm L_2}$ for unstable mass transfer through overfilling the outer Lagrangian surface of most RGB and AGB stars are around 1.4 to 2.0 (mass larger than $0.36~M_\odot$). Because the TPAGB stars have a very extended envelope and a comparable dynamical and thermal timescales, these donor stars very easily overfill their outer Lagrangian surfaces. So this kind of instability, through overfilling outer Lagrangian surface ($q_{\rm L_2}$, which is smaller than both $q_{\rm ad}$ and $q_{\rm th}$), dominates the late TPAGB donor stars with masses from $1.0$ to $5.0~M_\odot$.

\section{Discussion}

This study attempts to systematically survey the thresholds for thermal timescale mass transfer over the entire span of possible donor star evolutionary states. These thresholds mark bifurcation points close binary evolution, separating evolutionary channels proceeding on a thermal timescale (or unstable mass transfer through overfilling the outer Lagrangian point to form a CE) from those proceeding on a far more rapid timescale leading to CE evolution. 

The advantage of the thermal equilibrium mass loss model is that the donor star's response is independent of binary orbital evolution caused by mass or angular momentum loss from the binary system. We also discuss the shortcomings of our model briefly. The details of the subsequent process after the thermal timescale mass transfer are not studied here. The observed binary systems might have evolved away from the states after the mass transfer process finished. We assume the companion accretes all the material from its donor during thermal equilibrium mass transfer. However, the accretion process of the companion, and hence the response of the companion, has to be studied at the same time with the donor's mass transfer/loss process. What is more, the thermal timescales of the late RGB/AGB stars are typically only two orders of magnitude or so shorter than nuclear timescales. Throughout our simulations, we suppress composition changes along with our thermal equilibrium mass loss sequences, but still allow the convective boundaries to move. We adopt this treatment, which disregards the precise treatment of convection, since failing to suppress composition changes leads to numerical diffusion, which results in the illusion that nuclear fuel is drawn into burning shells, but is not consumed. Our treatment, while unphysical, is not expected to result in too great an aberration if the convective boundary is moving through a region of uniform composition. We check that this is indeed the case for most of our simulations.

Although we have the theoretical thresholds for a thermal timescale mass transfer or unstable mass transfer to form a CE in our series of papers, we are still missing the outcome of CE evolution. With the total energy as a function of the remaining mass in the adiabatic mass loss process, we can combine energy constraints with the requirements that both binary components fit within their post-common envelope Roche lobes. Then we can place strict limits on the masses, mass ratios, and remnant orbital separations of binaries passing through a common envelope evolution. We will present these results in the next paper of this series of rapid mass transfer in binaries.

\section{Summary}

In this paper, we present the thermal equilibrium mass loss model, which assumes the time derivative of the specific entropy, ${\rm d}s/{\rm d}t$, is frozen with mass. We have constructed model sequences describing the asymptotic responses of stars to mass loss in a binary system in thermal equilibrium, in which mass transfer is not so rapid that thermal relaxation is allowed within the stellar interior during mass loss. We assume an isothermal flow, which is transferred to its companion through ${\rm L_1}$ during thermal equilibrium mass transfer, can be accelerated to a thermal timescale mass transfer rate to get the inner radius, $R_{\rm KH}$. We compare the donor star's inner radius response with its Roche lobe radius response to get the critical mass ratio for thermal timescale mass transfer, $q_{\rm th}$. We also solve the critical mass ratio for overfilling the outer Lagrangian surface, $q_{\rm L_2}$, to be that initial mass ratio such that the volume-equivalent radius of the outer lobe, $R_{\rm L_2}$, is just tangent to the stellar radius, $R$, at one point along a mass-loss sequence. Using several $3.2~M_\odot$ stellar models as examples, we study the stellar response to thermal equilibrium mass loss and present the thresholds for thermal timescale mass transfer and unstable mass transfer through overfilling the outer Lagrangian point. The results show that, a binary system containing a late RGB/AGB donor star with an initial mass ratio larger than $q_{\rm L_2}$ could suffer an unstable mass transfer through the outer Lagrange point and might also result in a kind of CE evolution. 

We present the initial structure paramters and the thresholds for thermal time scale mass transfer with a stellar model grid that covers a full range of masses and evolutionary stages. Combining with the results in~\citet{ge15,ge19}, we can use this study as an input to population synthesis studies of close binary evolution that seek to quantify the frequency and properties of various possible evolutionary channels. The application in population synthesis studies might help us to explain the evolution channels of many related binary objects.

\acknowledgments

We are indebted to the referee, Christopher Tout, for addressing many improvements and a constructive review of this paper. This work was supported by grants from the National Natural Science Foundation of China (NSFC Nos. 11673058, 11733008, 11521303), the key research program of frontier sciences, CAS, NO. ZDBS-LY-7005, the Natural Science Foundation of Yunnan Province (Grant No. 2019HA012), and the Department of Astronomy at the University of Illinois at Urbana-Champaign. HG thanks the Chinese Academy of Sciences and the Department of Astronomy, the University of Illinois at Urbana-Champaign, for the one-year visiting program. HG thanks Prof. Xuefei Chen for helpful discussion on this work. HG also thanks Matthias Kruckow for improving the manuscript. RFW was supported in part by US National Science Foundation grants AST 04-06726 and AST 14-13367.



\end{document}